\documentclass[pra,twocolumn,showpacs,superscriptaddress]{revtex4-1}

\usepackage{epsf,epsfig}
\usepackage[psamsfonts]{amssymb}
\usepackage{amsmath}
\usepackage{bm}
\usepackage{natbib}
\usepackage{graphicx}
\usepackage{color}
\usepackage{mathrsfs}
\newcommand{\G}{\mbox{\sffamily\bfseries{G}}}

\begin{document}
\title{Quantum light propagation in a uniformly moving dissipative slab}
\author{Marzye Hoseinzadeh}
\affiliation{Department of Physics, Faculty of Basic Sciences, Shahrekord University, P.O. Box 115, Shahrekord 88186-34141, Iran.}
\affiliation{Department of Physics, Faculty of Science, University of Isfahan, Hezar Jerib, Isfahan, 81746-73441, Iran. }
\author{Ehsan Amooghorban}
\email{ehsan.amooghorban@sku.ac.ir}
\affiliation{Department of Physics, Faculty of Basic Sciences, Shahrekord University, P.O. Box 115, Shahrekord 88186-34141, Iran.}
\affiliation{Photonics Research Group, Shahrekord University, P.O. Box 115, Shahrekord 88186-34141, Iran.}
\affiliation{Nanotechnology Research Center, Shahrekord University, Shahrekord 88186-34141, Iran.}
\author{ Ali Mahdifar}
\affiliation{Department of Physics, Faculty of Science, University of Isfahan, Hezar Jerib, Isfahan, 81746-73441, Iran. }
\author{ Maryam Aghabozorgi Nafchi}
\affiliation{Department of Electrical Engineering, Faculty of Engineering and Computer Science, Shahrekord University, Shahrekord 88186-34141, Iran.}

\begin{abstract}
Within the framework of a phenomenological quantization scheme, we present the quantization of the electromagnetic field in the presence of a moving absorptive and dispersive magneto-dielectric slab (MDS) with uniform velocity in the direction parallel to its interface. We derive the quantum input-output relations for the case that quantum states propagate perpendicularly to the moving MDS.
We thoroughly investigate the impact of the motion of the moving MDS on quantum properties of the incident states. To illustrate this, by modeling the dispersive and dissipative effects of the slab by the Lorentz model in its rest
frame, we compute the quadrature squeezing and the Mandel parameter for the transmitted state when the incident states from left and right sides are, respectively, the coherent and the quantum vacuum states. It is shown that how quantum features of the incident state are
degraded when it is transmitted through the moving MDS.
\end{abstract}


\date{\today}
\maketitle
\section{Introduction}
The electrodynamics of moving media is one of the fundamental and old issues dates back as far as to the 1908, when Minkowski presented a covariant theory of electrodynamics in moving media by using Einstein's special relativity theory~\cite{Minkowski1908,Sommerfeld1964,Landau1984}. In particular, he introduced the relativistic constitutive equations to explain the electromagnetic phenomena related to the moving media.
Since then, the electrodynamics of moving media has been the subject of numerous theoretical and experimental investigations and applied in a variety of physical fields such as the optics of moving media~\cite{Landau1984,Gordon1923,Leonhardt1999,Carusotto2001,Carusotto2003,Strekalov2004}, radiation of fast charged particles in media~\cite{Nag1956}, and astrophysics~\cite{Kong1975}.

Among these studies, the problem of the scattering of electromagnetic waves from moving media is
a problem of both fundamental interest and practical importance which has long received much attention.
At first, Pauli and Sommerfeld were studied the frequency
shift of a reflected plane wave by a moving mirror~\cite{Pauli1958,Sommerfeld1959}.
Later on, many authors analyzed in detail this problem for the case of a moving half-space dielectric and moving dielectric
slab, and then generalized these calculations to an arbitrary direction of motion~\cite{Yeh1965,Yeh1966a,Yeh1966b,Payati1967,Shiozawa1967,Kong1968,Shiozawa1972,Kong1975,Huang1994}.

There has been a revival of interest in the scattering of electromagnetic waves by moving media over the past decade and has now become a rather topical area of research. Most of these works deal with fascinating and fundamental issues in the classical theory of electrodynamics
such as the analogy between light propagating in moving media and in curved
space-times~\cite{Leonhardt2000,Fiurasek2002}, the optical analogue of the Aharonov-Bohm effect~\cite{Cook1995}, the generation of negative refraction~\cite{Grzegorczyk2006}, the optically induced magnetoelectric effect~\cite{Da2007}, the occurrence of linear birefringence~\cite{Fumeron2011,Lin2016} and generation of coherent light by a moving medium~\cite{Svidzinsky2017}.
On the other hand, there are some problems such as the emission of light due to fast changes of the geometry~\cite{Dodonov2010,Silveirinha2014}, uniformly motion of media~\cite{Fulling1976,Barton1996,Volokitin2007,Maghrebi2012,Maghrebi2013} and rotating objects~\cite{Maghrebi2012,Maghrebi2013,Manjavacas2010,Zhao2012} which can only be illuminated in the framework of a full quantum approach.

The scattering of quantized waves from a dispersive and absorptive medium at rest has been discussed in Refs.~\cite{Gruner1996,Artoni1997,Artoni1998a,Artoni1998b,Matloob2000,Khanbekyan2003,Amooghorban2013,Amooghorban arXiv,Amooghorban2014,Aghbolaghi2017}, but to the best of our knowledge, this problem has not been studied before for the case of a moving medium whose velocity may have any value up to the speed of light in free space.
In this contribution, we intend to discuss such scattering behavior in detail. For an incident light with a nonclassical nature, the motion of the media definitely affects on quantum statistical properties of the light and put forwards some interesting results.
In order to assess these motion effects, one needs a quantum treatment for the propagation of electromagnetic fields in moving media.
To this end, we first have to quantize the electromagnetic field in the presence of an absorptive and dispersive moving medium. This is a more complicated task than the case of stationary media. Nevertheless, two phenomenological and canonical quantization schemes have been developed in unbounded moving media~\cite{Matloob2005a,Matloob2005b,Amooshahi2009,Kheirandish2011,Horsley2012}.

The use of sophisticated canonical approach, although strictly rigorous, may tend to obscure the simple physical concepts. Instead, we follow the phenomenological quantization scheme presented in~\cite{Matloob2005a,Matloob2005b} and extend this method to the more general and practical case in which a MDS surrounded by free space is in motion. It makes possible to investigate the scattering of an incident nonclassical light field by a moving polarizable and magnetizable slab in the laboratory, and we would expect the optical properties of a light propagating through a moving slab modify differently from the light propagating through the same slab at rest. In the quantization process, it is found that the motion leads to effective velocity-dependent electric permittivity and magnetic permeability which exhibit an anisotropic character for an observer in the laboratory frame, even if the medium is isotropic in its rest frame.
These effective parameters enable us to reduce the phenomenological scheme of the electromagnetic field quantization in a moving slab to a stationary slab in a straightforward manner.

The paper is organized as follows: In Sec.~\ref{Sec:Quantization}, we present the phenomenological quantization of the electromagnetic field in the presence of a moving MDS. We derive the input-output relations for quantized electromagnetic waves normally incident upon a MDS moving with uniform velocity in the direction parallel to its interface, and further analyze the reflection and
transmission coefficients of the moving MDS. In Sec.~\ref{Numerical Calculation and Analysis}, we study the effects of transmission of the coherent state (CS) through a moving slab on some statistical properties, such as quadrature squeezing and photon counting statistics.
A summary and interesting conclusions are deduced in Sec.~\ref{Sec:summary}. We provide further details of our calculations for the square root of the imaginary
part of effective  tensors, boundary conditions, the elements of transformation and absorbing matrices in Appendices~\ref{App:Square root of tensors},~\ref{App:Boundary Conditions},~\ref{App:Components of Transformation Matrix} and~\ref{App:Components of Absorbing Matrix}.

\section{Quantization of the electromagnetic field in the presence of a moving MDS}\label{Sec:Quantization}
\subsection{Basic equations}
%
Consider a homogeneous isotropic MDS moving uniformly at velocity ${\bf v}$ with respect to the laboratory frame. In the rest frame of the medium, the electromagnetic response of the slab is characterized by the electric permittivity $\varepsilon $
and magnetic permeability $\mu $ that satisfy the Kramers-Kronig relations. We assume that the moving MDS is surrounded by the vacuum, i.e., the permittivity and permeability in
the regions outside the moving MDS is unity.
The propagation of the classical electromagnetic waves in this slab can be described by the macroscopic Maxwell equations together with suitable constitutive relations.
From the point of view of a reference system comoving with the medium, the macroscopic Maxwell equations have the same mathematical form to that of an observer in the laboratory frame~\cite{Landau1984}.
Whereas, due to the uniform motion of the medium,
the relativistic relations between the  macroscopic electromagnetic fields ${\bf E}$, ${\bf D}$, ${\bf B}$ and ${\bf H}$ in the laboratory frame are given by the well known Minkowski constitutive relations.
Taking the Fourier transform of these fields with respect to the time, we can write the Minkowski constitutive relations for the positive frequency
part of the fields as~\cite{Chen1983}:
\begin{subequations}\label{Minkowski equations}
\begin{eqnarray}
{\bf{D}}^{+}({{\bf r},\omega})&=&{{\varepsilon }_{0}}\varepsilon {\bar{\bar{\boldsymbol\alpha }}}\cdot{\bf{E}}^{+}({{\bf r},\omega})+\frac{1}{c} {\bf \bar{\bar{m}}} \cdot
{\bf{H}}^{+}({{\bf r},\omega}), \,\,\,\,\,\,\,\,\,\,\label{Minkowski equations D}
\\
{\bf{B}}^+({{\bf r},\omega})&=&-\frac{1}{c}{\bf{\bar{\bar{m}}}}\cdot{\bf{E}}^+({{\bf r},\omega})+{{\mu }_{0}}\mu {\bar{\bar{\boldsymbol\alpha }}}\cdot{\bf{H}}^+({{\bf r},\omega}),\label{Minkowski equations H}\,\,\,\,\,\,\,\,\,\,
\end{eqnarray}
\end{subequations}
where boldface symbols and boldface symbols along with
double overlines are used to identify vector and second-rank tensor quantities, respectively.
Here, ${\bar{\bar{\rm\bf I}}}$ is the unit tensor, ${\bf \bar{\bar{m}}}$ is an antisymmetric tensor defined as ${\bf{m}}\times {\bar{\bar{\rm\bf I}}}$, and the symmetric tensor ${{\bar{\bar{\boldsymbol\alpha }}}}$ and the vector ${\bf m}$ are defined as:
\begin{equation}\label{symmetric tensor}
{{\bar{\bar{\boldsymbol\alpha }}}}= {\bf{\bar{\bar{I}}}}\alpha+\left( 1-\alpha  \right){\hat{v}\hat{v}} ,
\end{equation}
\begin{equation}\label{vector m}
{\bf m}=m{\hat{v}},
\end{equation}
where $m={\beta ( {{n}^{2}}-1 )}/({1-{{n}^{2}}{{\beta }^{2}}})$, $\beta ={v}/{c}$ and $\alpha ={\left( 1-{{\beta }^{2}} \right)}/{\left( 1-{{n}^{2}}{{\beta }^{2}} \right)}$ with $n=\sqrt{\varepsilon \mu }$ is refractive index of the medium in the rest frame of the medium, $\hat{v}$ is the unit vector of the velocity of the medium, and $c$ the
velocity of light in free space.

As mentioned in introduction, the phenomenological quantization of the
electromagnetic fields in unbounded moving media has been developed previously~\cite{Matloob2005a,Matloob2005b}.
Here, we briefly summarized the main results needed for an understanding of the present paper, and then extend this approach to the practical case of the moving MDS.

Unfortunately, the above form of the constitutive relations are not convenient
for the phenomenological approach. Because, unlike to the electromagnetic fields, the electric permittivity and the
magnetic permeability are the parameters observed in the
rest frame of the medium. Using the Maxwell equations in reciprocal space, these constitutive relations can be cast into more convenient form as~\cite{Matloob2005a,Matloob2005b}:
\begin{subequations}
\begin{eqnarray}
{\bf{D}}^{+}({{\bf k},\omega})={{\varepsilon }_{0}}{{\bar{\bar{\boldsymbol\varepsilon }}}_{\rm eff}}\cdot {\bf{E}}^{+}({{\bf k},\omega})+{{\bf{P}}^{+}_{N}}({{\bf k},\omega}),\label{new constructionist relations E}
\\
{\bf{H}}^+({{\bf k},\omega})={{\mu }^{-1}_{0}}{{\bar{\bar{\boldsymbol\mu }}}^{-1}_{\rm eff}}\cdot{\bf{B}}^+({{\bf k},\omega})-{{\bf{M}}^+_{N}}({{\bf k},\omega}),\label{new constructionist relations H}
\end{eqnarray}
\end{subequations}
where ${{{{\boldsymbol\varepsilon }}}_{\rm eff}}$ and ${{{{\boldsymbol\mu }}}_{\rm eff}}$ are the effective electric permittivity and
the magnetic permeability tensors which show the material parameters of the moving medium as seen by the observer in the laboratory frame. The explicit form of these nonsymmetric tensors are given by~\cite{Matloob2005a,Matloob2005b}
\begin{subequations}\label{effective material tensors}
\begin{eqnarray}
{{\bar{\bar{\boldsymbol\varepsilon }}}_{\rm eff}}=\varepsilon \,\left( {\bar{\bar{\boldsymbol\alpha }}}+\frac{({\bar{\bar{\boldsymbol\alpha }}}\cdot
{\bf{P}}) {\bf{m}}-({\bf{m}}\cdot {\bf{P}})\,{\bar{\bar{\boldsymbol\alpha }}}}{{{\alpha }^{2}}{{n}^{2}}} \right),\label{effective electric permittivity tensor}
\\
{{\bar{\bar{\boldsymbol\mu }}}_{\rm eff}}=\mu \,\left( {\bar{\bar{\boldsymbol\alpha }}}+\frac{({\bar{\bar{\boldsymbol\alpha }}}\cdot {\bf{P}})\,{\bf{m}}-({\bf{m}}\cdot
{\bf{P}})\,{\bar{\bar{\boldsymbol\alpha }}}}{{{\alpha }^{2}}{{n}^{2}}} \right),\label{effective magnetic permeability tensor}
\end{eqnarray}
\end{subequations}
where ${{\bf{P}}={q^{-1}}{\bf{k}}+{\bf{m}}}$ in which $q={\omega }/{c}$ and ${\bf{k}}$ is the wave vector.
From above equations, one observes that the field ${\bf D}$(${\bf B}$) is not in the direction of ${\bf E}$(${\bf H}$); hence a moving medium is anisotropic even thought it is isotropic in the rest frame of the the medium.
Furthermore, it is seen that the noise polarization ${{\bf{P}}_{N}}$ and the noise magnetization ${{\bf{M}}_{N}}$
are added to the Minkowski constitutive relations~(\ref{Minkowski equations}) due to material absorption associated with
the electric and magnetic losses.
These noise operators are unavoidably required in order to preserve the well-known canonical field commutation relations, and therefore, lead to the correct Heisenberg equations of motion for the fields~\cite{Matloob2005a,Matloob1996,Knoll2001}. In this manner, the quantization of electromagnetic fields in presence of moving slab can be carried out similar to the method has been accomplished for a stationary anisotropic slab in Refs.~\cite{Dong2011,Hoseinzadeh2017}.

By making use of the Weyl gauge, which has the advantage that the scalar potential is zero, and by combining the Maxwell equation and Minkowski constitutive relations, the wave equation for the the positive frequency part of the vector potential operator is obtained as:
\begin{eqnarray}\label{wave equation for vector potential operator}
\Big({ {{\bf \bar{\bar{k}}}  \cdot {{{{\bar{\bar{\boldsymbol \mu }}}}}^{-1}_{\rm eff}} \cdot {\bf \bar{\bar{k}} }} +{q^2} {{{\bar{\bar{\boldsymbol\varepsilon }}}}_{\rm eff}}}\Big) \cdot {\bf \hat{A}}^+({\bf k},\omega)=-\mu_0 {{{{\hat{\bf{J}}}}}^{+}_{N}}( {\bf k},\omega )
\end{eqnarray}
where ${{\bf \bar{\bar{k}}}}= {\bf k}\times {\bar{\bar{\rm\bf I}}}$ is an antisymmetric tensor and ${{{{\hat{\bf{J}}}}}^{+}_{N}}( {\bf k},\omega )=-i\omega {{{\bf{P}}}^{+}_{N}}( {\bf k},\omega )+ i{{\bf \bar{\bar{k}}}} \cdot {{{\bf{M}}}^{+}_{N}}( {\bf k},\omega )$ is the Fourier-transformed current operator.
Neglect of this noise operator leads to a spatially damped vector potential, and therefore, the canonical field commutation relations
are not preserved.
This noise operator operator can be described in term of the fundamental variables of the system as follows~\cite{Matloob2005a}:
\begin{eqnarray}\label{current density}
 {{{{\hat{\bf{J}}}}}_N^+}\left( {\bf k},\omega  \right)
  &=& \omega \sqrt{\frac{2\hbar \,
 {{\varepsilon }_{0}}}{ S}{\bar{\bar{\boldsymbol\varepsilon }}}_{\rm eff}^{I}}\cdot {{{{\hat{\bf{f}}}}}_{e}}\left( {\bf k},\omega  \right)\\
&&+ i{{\bf \bar{\bar{k}}}} \cdot \sqrt{\frac{-2\hbar }{ S{{\mu }_{0}}}{{{\bar{\bar{\boldsymbol\mu }}}}^{-1\,{I}}_{\rm eff}}}\cdot {{{{\hat{\bf{f}}}}}_{m}}\left( {\bf k},\omega
 \right),\nonumber
\end{eqnarray}
where the superscript $I$ stand for the imaginary part of a function, $S$ is the transverse normalized area, and $\hat{{\bf f}}_{e}({\bf k},\omega)$ and $\hat{{\bf f}}_{m}({\bf k},\omega)$ are the Fourier transforms of the bosonic field operators
$\hat{{\bf f}}_{e}({\bf r},\omega)$ and $\hat{{\bf f}}_{m}({\bf r},\omega)$ for the electric and magnetic excitations of the system.
The components of these bosonic operators satisfy the commutation relations:
\begin{subequations}
\begin{eqnarray}
{[\hat{f}_{\lambda,i}({\bf k},\omega),\hat{f}^\dagger_{\lambda',j}({\bf k}',\omega')]}&=&{\delta_{\lambda \lambda'}\delta_{ij}\delta({\bf k}-{\bf k}')\delta(\omega-\omega')},\,\,\,\,\,\,\,\,
\\
{[\hat{f}_{\lambda,i}({\bf k},\omega),\hat{f}^\dagger_{\lambda',j}({\bf k}',\omega')]}&=&{ 0},
\end{eqnarray}
\end{subequations}
where $\lambda,\lambda'=e,m$ and $i,j=x,y,z$.
Moreover, in writing Eq.~(\ref{current density}), the square root of the imaginary
part of the effective optical tensors are defined as~\cite{Matloob2005a}
\begin{subequations}
\begin{eqnarray}\label{square root imaginary tensors}
\sqrt{{\bar{\bar{\boldsymbol\varepsilon }}}_{\rm eff}^{I}}\cdot \sqrt{{\bar{\bar{\boldsymbol\varepsilon }}}_{\rm eff}^{I}}^\dag &=&\frac{i}{2}\Big({\bar{\bar{\boldsymbol\varepsilon }}}_{\rm eff}-{\bar{\bar{\boldsymbol\varepsilon }}}_{\rm eff}^\dag \Big)^*,\\
\sqrt{{{{\bar{\bar{\boldsymbol\mu }}}}^{-1\,{I}}_{\rm eff}}} \cdot \sqrt{{{{\bar{\bar{\boldsymbol\mu }}}}^{-1\,{I}}_{\rm eff}}}^\dag &=&\frac{i}{2}
\Big({{{\bar{\bar{\boldsymbol\mu }}}}^{-1}_{\rm eff}}-{{{\bar{\bar{\boldsymbol\mu }}}}^{-1}_{\rm eff}}^\dag \Big)^*.
\end{eqnarray}
\end{subequations}

With the help of Eq.~(\ref{wave equation for vector potential operator}), and by using the inverse Fourier transforms, the explicit form of the vector potential operator in the real space can be obtained as
\begin{eqnarray}\label{vector potential}
{{\bf\hat{A}}}\left( {\bf r},\omega \right)&=&-\frac{{{\mu }_{0}}}{4\pi^2} \int_0^\infty d\omega \int d^3{\bf k} \\
&&\times\Big(\bar{\bar\G}( {\bf k},\omega )\cdot {{{{\bf{J}}}}_{N}}( {\bf k},\omega )e^{i({\bf k}\cdot {\bf r}- \omega t)}+{\rm H.C.}\Big),\nonumber
\end{eqnarray}
where $ \bar{\bar\G} $ is the electromagnetic Green's tensor (GT) of the system as uniquely defined by
the Helmholtz equation: $\Big({ \bar{\bar{\bf k}} } \cdot {{\bar{\bar{\boldsymbol{\mu} }}}}^{-1}_{\rm eff} \cdot {\bf \bar{\bar{k}}} +q^2 {{{\bar{\bar{\boldsymbol{\varepsilon} }}}}_{\rm eff}}\Big) \cdot \bar{\bar\G}( {\bf k},\omega )=\bar{\bar{{\bf I}}}$, subject to the appropriate boundary condition.
It is known that the solution of this Helmholtz equation for an infinite moving medium is given by~\cite{Matloob2005a}:
\begin{equation}\label{Green tensor}
\bar{\bar\G}\left( {\bf k},\omega  \right)=\frac{1}{{{q}^{2}}\varepsilon }\frac{\left( {\bf k}+q {\bf m} \right)\left( {\bf k}+q {\bf m}
\right)-{{n}^{2}}{{q}^{2}}{{\alpha }^{2}}{{{\bar{\bar{\boldsymbol\alpha }}}}^{-1}}}{\left( {\bf k}+q {\bf m} \right)\cdot \bar{\bar{\boldsymbol\alpha }}\cdot \left(  {\bf k}+q {\bf m}
\right)-{{n}^{2}}{{q}^{2}}{{\alpha }^{2}}}.
\end{equation}
Let us consider the $z$ axis of the laboratory Cartesian frame is normal to the interface of the slab that moving
with a constant velocity in the direction parallel to its interface.
Interfaces are in the $xy$ plane, and the coordinate origin is taken at the center of the slab.
Without loss of the generality, we assume that the slab is moving with a constant speed along the $y$ axis, {\rm i.e.},  $\text{\textbf{v}=}{{v}}\text{\^{y}}$, as illustrated schematically in Fig.~\ref{Fig:MDS slab}.
By recalling Eq.~(\ref{effective material tensors}), the effective tensors ${{{\bar{\bar{\boldsymbol\varepsilon }}}}_{\rm eff}}$ and ${{{\bar{\bar{\boldsymbol\mu }}}}_{\rm eff}}$ of the moving MDS are simplified as:
\begin{subequations}\label{effective permittivity and permeability material tensor}
\begin{eqnarray}
{{{\bar{\bar{\boldsymbol\varepsilon }}}}_{\rm eff}}=\varepsilon \left( \begin{matrix}
   \frac{{{\alpha }^{2}}{{n}^{2}}-{{m}^{2}}}{\alpha {{n}^{2}}} & 0 & 0  \\
   0 & 1 & 0  \\
   0 & \frac{c{{k}_{z}}m}{\omega \alpha {{n}^{2}}} & \frac{{{\alpha }^{2}}{{n}^{2}}-{{m}^{2}}}{\alpha {{n}^{2}}}  \\
\end{matrix} \right),\label{effective electric permittivity material}
\\
{{{\bar{\bar{\boldsymbol\mu }}}}_{\rm eff}}=\mu \left( \begin{matrix}
   \frac{{{\alpha }^{2}}{{n}^{2}}-{{m}^{2}}}{\alpha {{n}^{2}}} & 0 & 0  \\
   0 & 1 & 0  \\
   0 & \frac{c{{k}_{z}}m}{\omega \alpha {{n}^{2}}} & \frac{{{\alpha }^{2}}{{n}^{2}}-{{m}^{2}}}{\alpha {{n}^{2}}}  \\
\end{matrix} \right).\label{effective magnetic permeability tensor material}
\end{eqnarray}
\end{subequations}
Here, for the sake of convenience, the effective tensors are written in matrix
form. Accordingly, after lengthy
calculations, the square root of the imaginary part of the effective tensors $\sqrt{{\bar{\bar{\boldsymbol\varepsilon }}}_{\rm eff}^{I}\left( \omega  \right)}$ and $\sqrt{{\bar{\bar{\boldsymbol\mu }}}_{\rm eff}^{-1\,I}\left( \omega  \right)}$ can be obtained as:
\begin{subequations}
\begin{eqnarray}
\sqrt{{\bar{\bar{\boldsymbol\varepsilon }}}_{\rm eff}^{I}\left( \omega  \right)}&=&\left( \begin{matrix}
   e_{11} & 0 & 0  \\
   0 & e_{22} & e_{23}  \\
   0 & e_{32} & e_{33} \\
\end{matrix} \right) , \label{the square root of the imaginary part of effectively permittivity}
\\
\sqrt{{\bar{\bar{\boldsymbol\mu }}}_{\rm eff}^{-1\,I}\left( \omega  \right)}&=&\left( \begin{matrix}
   m_{11} & 0 & 0  \\
   0 & m_{22} &  m_{23}  \\
   0 & m_{32} & m_{33}  \\
\end{matrix} \right),\label{the square root of the imaginary part of effectively magnetic permeability}
\end{eqnarray}
\end{subequations}
where the explicit form of these elements has been presented in Appendix~\ref{App:Square root of tensors}.
From here on, we restrict our attention to the case that a quantized linearly polarized wave normally incident along $z$ axis toward the moving slab.
Using Eq.~(\ref{Green tensor}) and the fact that the slab is translationally invariant in the plane of the surface, the GT of system in the reciprocal space is obtained as:
\begin{eqnarray}\label{Green tensor of  in frequency-domain and wave vector}
\bar{\bar\G}\left( {\bf k},\omega  \right)&=&\frac{1}{{{q}^{2}}\varepsilon \left( k_{z}^{2}\alpha +{{q}^{2}}{{m}^{2}}-{{n}^{2}}{{q}^{2}}{{\alpha }^{2}}
\right)}\\
&&\times\left( \begin{matrix}
   -{{n}^{2}}{{q}^{2}}\alpha  & 0 & 0  \\
   0 & {{q}^{2}}{{m}^{2}}-{{n}^{2}}{{q}^{2}}{{\alpha }^{2}} & qm{{k}_{z}}  \\
   0 & qm{{k}_{z}} & k_{z}^{2}-{{n}^{2}}{{q}^{2}}{{\alpha }^{2}}  \\
\end{matrix} \right).\nonumber
\end{eqnarray}
By using the inverse Fourier transformation and taking the contour integration over $k_z$, we straightforwardly arrive at the electromagnetic GT in the coordinate space as follows:
\begin{eqnarray}\label{Green tensor system in space of coordinates}
\bar{\bar\G}\left( z,{z}',\omega  \right)=
 e^{ik\left| z-{z}' \right|}  \left({ \begin{matrix}
   \frac{-i\mu }{2k} & 0 & 0  \\
   0 & \frac{-i{{\mu }_{\rm eff ,xx}}}{2k} & \frac{im}{2\alpha \varepsilon {\omega }/{c}}  \\
   0 & \frac{im}{2\alpha \varepsilon {\omega }/{c}} & \frac{i\left( k^2c^2/\omega^2-{{n}^{2}}{{\alpha }^{2}} \right)}{2\alpha \varepsilon
   k}  \\
\end{matrix} }\right). \nonumber\\
\end{eqnarray}
Here, $k=n_{\rm eff}{\omega }/{c}$ where $n_{\rm eff}=\sqrt{({{{\alpha }^{2}}{{n}^{2}}-{{m}^{2}}})/{\alpha }}={\gamma
} \sqrt{{{n}^{2}}-\beta^2} $ is the refraction index in the laboratory frame, in which $\gamma=(1-\beta^2)^{-1/2} $ is the usual Lorentz factor. This refraction index is in agreement with the classical results derived earlier in a different way for dispersion relation of moving media~\cite{Chen1983}.

By combining Eqs.~(\ref{Green tensor system in space of coordinates}) and~(\ref{vector potential}), the vector potentials for different polarized waves can be represented in a convenient form as:
\begin{subequations}\label{vector potential components xy}
\begin{eqnarray}
\hat{A}_{x}\left( z,\omega \right)&=&\sqrt{\frac{\hbar \xi \left( \omega \right)}{32\pi^4 S{{\varepsilon }_{0}}c\,\omega}}\frac{\mu }{n_{\rm eff}}\left[{{{e}^{{i\eta
\omega z}/{c}}}{{\hat{a}}_{x+}}\left( z,\omega \right)}\right.\nonumber\\
&&\left.{+{{e}^{{-i\eta \omega z}/{c}}}{{\hat{a}}_{x-}}\left( z,\omega \right)+H.C.}\right],\label{vector potential components x}
\\
\hat{A}_{y}\left( z,\omega  \right)&=&\sqrt{\frac{\hbar {\xi }'\left( \omega  \right)}{32\pi^4 S{{\varepsilon }_{0}}c\,\omega }}\frac{\mu_{{\rm
eff ,xx}}}{n_{\rm eff}}\left[{{{e}^{{i\eta \omega z}/{c}}}{{\hat{a}}_{y+}}\left( z,\omega  \right)}\right.\nonumber \\
&&\left.{+{{e}^{{-i\eta \omega z}/{c}}}{{\hat{a}}_{y-}}\left(
z,\omega  \right)+H.C.}\right],\label{vector potential components y}
\end{eqnarray}
\end{subequations}
where the operators
\begin{subequations}\label{annihilation bosonic operators}
\begin{eqnarray}
&&{{{\hat{a}}}_{x\pm }}\left( z,\omega  \right)=i\sqrt{2\kappa \left( \omega  \right){\omega }/{c}}\,{{e}^{\mp \kappa ( \omega )z{\omega
  }/{c} }}\nonumber\\
&&\times\int_{-\infty }^{\pm z}{dz'{e^{-i n_{\rm eff}( \omega )z'{\omega }/{c}}}}\left[{\frac{\sqrt{\bar{\bar\varepsilon }_{\rm
  eff ,xx}^{I}}{{{\hat{f}}}_{e,x}}\left( \pm \,{z}',\omega  \right) }{\sqrt{\bar{\bar\varepsilon}_{\rm eff ,xx}^{I}+{{\left| n_{\rm eff} \left( \omega  \right) \right|}^{2}} E_m  } } }\right.\nonumber\\
&&\,\,\,\,\,\,\,\,\,\, \pm \left. \frac{   n_{\rm eff}\left( \omega  \right)\sqrt{E_m} {{\hat{f}}_{m,\bot }}\left( \pm z',\omega  \right) }{\sqrt{\bar{\bar\varepsilon}_{\rm eff ,xx}^{I}+{{\left| n_{\rm eff} \left( \omega  \right) \right|}^{2}} E_m }}\right],\label{annihilation operators x}
\\
\nonumber\\
&&{{{\hat{a}}}_{y\pm }}\left( z,\omega  \right)= i\sqrt{2\kappa \left( \omega  \right){\omega }/{c}}\,{{e}^{\mp \kappa \left( \omega  \right)z{\omega
  }/{c}}}\nonumber\\
&&\times\int_{-\infty }^{\pm z}{d\,{z}'\,{{e}^{-i n_{\rm eff}\left( \omega  \right){z}'{\omega }/{c}}}}
\left[{\frac{\sqrt{E_e}\,{{{\hat{f}}}_{ e,\bot}}\left( \pm {z}',\omega
  \right) }{\sqrt{ E_e-{{\left| n_{\rm eff}\left( \omega  \right) \right|}^{2}}\bar{\bar\mu}_{\rm eff ,xx}^{-1\,I}}}\nonumber }\right.\\
&&\,\,\,\,\,\,\,\,\pm \left.{\frac{  i n_{\rm eff}\left( \omega  \right)\sqrt{-\bar{\bar\mu}_{\rm eff ,xx}^{-1\,I}}\,\,{{{\hat{f}}}_{m,x}}\left( \pm {z}',\omega
 \right)}{\sqrt{ E_e -{{\left| n_{\rm eff}\left( \omega  \right) \right|}^{2}}\bar{\bar\mu}_{\rm eff ,xx}^{-1\,I}}} }\right].\label{annihilation operators y}
\end{eqnarray}
\end{subequations}
are associated with the amplitudes of the $x$ and $y$ polarized waves propagating to the right $(+)$ and
left $(-)$. Here, $\eta $ and $\kappa $ are, respectively, the real and imaginary parts of the refractive index $n_{\rm eff}$, and the parameters $\xi \left( \omega  \right)$ and ${\xi }'\left( \omega  \right)$ are, respectively, defined as ${(\bar{\bar\varepsilon}_{\rm eff ,xx}^{I}+{{\left|
n_{\rm eff}  \right|}^{2}}E_m})/{2\kappa }$
and ${( E_e -{{\left| n_{\rm eff} \right|}^{2}}\bar{\bar\mu}_{\rm eff ,xx}^{-1\,I})}/{2\kappa }$,
in which $E_e={{\left| e_{22}-\left( mn_{\rm eff}/\varepsilon \alpha {{\mu }_{\rm eff ,xx}} \right){{e}_{32}} \right|}^{2}}+{{\left| e_{23}-
mn_{\rm eff}{{e}_{33}/\varepsilon \alpha {{\mu }_{\rm eff ,xx}}} \right|}^{2}}$ and $E_m=|m_{22}|^2+|m_{23}|^2 $. Moreover, the new bosonic field operators ${{\hat{f}}_{e,\bot }}\left( z,\omega  \right)$ and ${{\hat{f}}_{m,\bot }}\left( z,\omega  \right)$ are defined as:
\begin{subequations}\label{new bosonic operator }
\begin{eqnarray}
{{\hat{f}}_{e,\bot }}\left( z,\omega  \right)&=&\frac{ \left( e_{22}-\frac{m\,n_{\rm eff}}{{\varepsilon} \alpha {{\mu}_{\rm
eff ,xx}}}{{e}_{32}} \right){{{\hat{f}}}_{e,y}}\left( z,\omega  \right)}{\sqrt{E_e}}\\
&&+\frac{\left( e_{23}-\frac{m\,n_{\rm eff}}{{\varepsilon} \alpha {{\mu}_{\rm
eff ,xx}}}{{e}_{33}} \right){{{\hat{f}}}_{e,z}}\left( z,\omega  \right)}{\sqrt{E_e}},\nonumber\\
{{\hat{f}}_{m,\bot }}\left( z,\omega  \right)&=& \frac{m_{22} {{{\hat{f}}}_{m,y}}\left( {z},\omega
 \right)+m_{23} {{{\hat{f}}}_{m,z}}\left( {z},\omega
 \right)}{\sqrt{E_m}},\,\,\,\,\,\,\,\,\,\,\,\,\,\,
\end{eqnarray}
\end{subequations}
which satisfy the following bosonic commutation relations:
\begin{subequations}\label{commutation relations of new bosonic operator}
\begin{eqnarray}
\left[ {{{\hat{f}}}_{\nu,\bot  }}\left( z,\omega  \right),\hat{f}_{\nu',\bot }^{\dagger }\left( {z}',{\omega }' \right) \right]&=&\delta_{\nu\nu'}\delta( z-z')\delta ( \omega -{\omega }' ) ,\,\,\,\,\,\,\,\,\,\,\,
\\
\left[ {{{\hat{f}}}_{\nu, \bot  }}\left( z,\omega  \right),{{{\hat{f}}}_{ \nu',\bot}}\left( {z}',{\omega }' \right) \right]&=&0,
\end{eqnarray}
\end{subequations}
where $\nu=e,m$. With these relations in mind,
the amplitude operators ${{\hat{a}}_{x\pm }}\left( z,\omega  \right)$ and ${{\hat{a}}_{y\pm }}\left( z,\omega  \right)$ are found to satisfy the commutation relations as follows:
\begin{subequations}
\begin{eqnarray}
\Big[ {{{\hat{a}}}_{\sigma \pm }}\left( z,\omega  \right),\hat{a}_{{\sigma }'\pm }^{\dagger }\left( {z}',{\omega }' \right) \Big]&=&{{\delta }_{\sigma {\sigma }'}}\delta \left( \omega -{\omega
}' \right),\,\,\,\,\,\,\,\,\,\,\,\,\,\label{commutation relations of annihilation operators}
\\
\Big[ {{{\hat{a}}}_{\sigma \pm }}\left( z,\omega  \right),{{{\hat{a}}}_{{\sigma }'\pm }}\left( {z}',{\omega }' \right) \Big]&=&0,\label{commutation relations of annihilation operators}
\end{eqnarray}
\end{subequations}
where $\sigma ,{\sigma }'=x,y$. From Eq.~(\ref{annihilation bosonic operators}), it can be easily shown that these operators satisfy the following quantum Langevin equations
%
\begin{subequations}\label{Langevin equations xy}
\begin{eqnarray}
\frac{\partial }{\partial z}{{\hat{a}}_{x\,\pm }}\left( z,\omega  \right)&=&\mp \kappa {\omega }/{c}{{\hat{a}}_{x\,\pm }}\left( z,\omega  \right)\pm {{\hat{D}}_{x\,\pm
}}\left( z,\omega  \right),\,\,\,\,\,\label{Langevin equations x}
\\
\frac{\partial }{\partial z}{{\hat{a}}_{y\pm }}\left( z,\omega  \right)&=&\mp \kappa {\omega }/{c}{{\hat{a}}_{y\pm }}\left( z,\omega  \right)\pm {{\hat{D}}_{y\pm
}}\left( z,\omega  \right),\label{Langevin equations y}
\end{eqnarray}
\end{subequations}
in which, the Langevin noise operators ${{\hat{D}}_{\sigma \pm }}\left( z,\omega  \right)$ are defined as:
\begin{subequations}\label{the relations of operators Dx Dy}
\begin{eqnarray}
&&\hspace{-0.1cm}{{\hat{D}}_{x\pm }}\left( z,\omega  \right)= \pm i\sqrt{2\kappa \left( \omega  \right){\omega }/{c}}\,{{e}^{\mp i \eta \left( \omega  \right)\omega
{z}/{c}}}\\
&&\times\frac{\sqrt{\varepsilon _{\rm eff ,xx}^{I}}\,{{{\hat{f}}}_{e,x}}\left( \pm {z}',\omega  \right)\pm  n_{\rm eff} ( \omega ) \sqrt{E_m} {{\hat{f}}_{m,\bot }}\left( \pm z',\omega  \right)}{\sqrt{\varepsilon _{\rm eff ,xx}^{I}+{{\left| n_{\rm eff} ( \omega  ) \right|}^{2}}E_m }},\label{the relations of operators Dx} \nonumber
\\
&&\hspace{-0.1cm}{{\hat{D}}_{y\,\pm }}\left( z,\omega  \right) = \pm i\sqrt{2\kappa \left( \omega  \right){\omega }/{c}}\,{{e}^{\mp i \eta \left( \omega  \right)\,\omega
{z}/{c}}}\\
&&\times\frac{\sqrt{E_e}{{{\hat{f}}}_{e,\bot}}\left( \pm {z}',\omega  \right)\pm i n_{\rm eff}\left( \omega  \right)\sqrt{-\mu_{\rm eff ,xx}^{-1\,I}}{{{\hat{f}}}_{m,x}}\left(
\pm {z}',\omega  \right)}{\sqrt{  E_e-{{\left| n_{\rm eff}\left( \omega  \right) \right|}^{2}}\mu_{\rm eff ,xx}^{-1\,I}}}.\label{the relations of operators Dy}\nonumber
\end{eqnarray}
\end{subequations}
The equations (\ref{Langevin equations xy})-(\ref{the relations of operators Dx Dy}) will make it possible to calculate
output fields in terms of input fields at any position outside
the moving slab, without explicitly using the Green function~(\ref{Green tensor system in space of coordinates}). We do this job in the next subsection.
%
\subsection{Quantum optical input-output relations for a moving MDS}
\begin{figure}[t]
\centering
\includegraphics[scale=0.35]{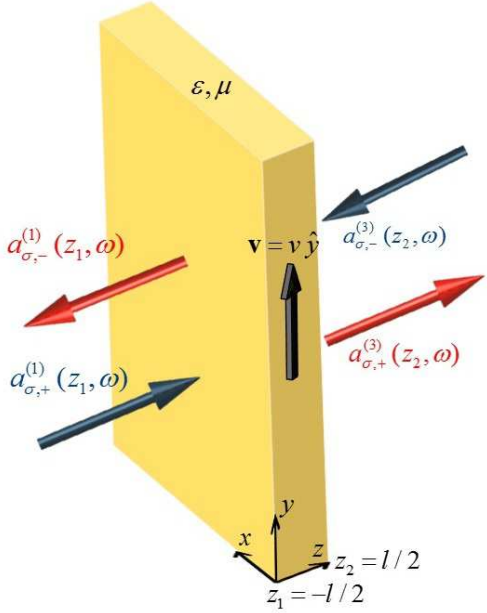}
\caption{ The geometry representation of the system for the fields impinging leftwards and rightwards on the MDS which is moving perpendicular to the incident field and parallel to its outer surface. The arrows together with the bosonic operators show the input and the output modes defined in the input-output relations~(\ref{input - output quantum relation}).}
\label{Fig:MDS slab}
\end{figure}

Let us consider the quantization method developed in the previous section for a magnetodielectric slab moving uniformly with velocity ${\bf v}=v \hat{y}$ parallel to its outer surface.
From the perspective of an observer in the laboratory frame and based on Eq.~(\ref{effective material tensors}), the optical properties of the slab can be described by the effective permittivity and permeability tensors~(\ref{effective electric permittivity material}) and~(\ref{effective magnetic permeability tensor material}).
As illustrated schematically in Fig.~\ref{Fig:MDS slab}, the magnetodielectric slab is taken to have a thickness $l$ with boundaries at ${{z}}=\pm{l}/{2}$. We introduce the annihilation operators of the incoming and the outgoing radiations on the left and the right sides of
the slab by $\hat{a}_{\sigma \pm }^{\left( 1 \right)}\left( z,\omega \right)$ and $\hat{a}_{\sigma \pm}^{\left( 3 \right)}\left( z,\omega  \right)$, respectively.

Let us now derive the input-output relations for the moving MDS, without explicitly applying
the GT~(\ref{Green tensor system in space of coordinates}). To accomplish this goal, we proceed in three steps:
First, by using Eqs.~(\ref{annihilation operators x}) and~(\ref{annihilation operators y}), we relate the amplitude operators $\hat{a}_{\sigma \pm }^{\left( 2 \right)}\left( z,\omega \right)$ within the slab at the positions $z = \pm l/2$ to each other as:
\begin{equation}\label{operators within slab}
\left( \begin{matrix}
   \hat{a}_{\sigma +}^{\left( 2 \right)}\left( {l}/{2},\omega  \right)  \\
   \hat{a}_{\sigma -}^{\left( 2 \right)}\left( {l}/{2},\omega  \right)  \\
\end{matrix} \right)={{\mathbb R}_{\sigma }}\left( \begin{matrix}
   \hat{a}_{\sigma +}^{\left( 2 \right)}\left( {-l}/{2},\omega  \right)  \\
   \hat{a}_{\sigma -}^{\left( 2 \right)}\left( {-l}/{2},\omega  \right)  \\
\end{matrix} \right)+\left( \begin{matrix}
   \hat{d}_{\sigma +}  \\
   \hat{d}_{\sigma -}  \\
\end{matrix} \right),
\end{equation}
where ${\sigma=x,y }$ denotes the polarization of different modes, and ${{\mathbb R}_{\sigma }}$ is a diagonal $2 \times 2$ matrix with ${{R}_{\sigma
,11}}={1}/{{{R}_{\sigma ,22}}}={{e}^{-\kappa \omega l/{c}}}$. The quantum noise operators in the
matrix equation~(\ref{operators within slab}) are given by:
\begin{equation}\label{operators D}
\hat{d}_{\sigma \pm }={{e}^{\mp {\kappa}\omega {l}/{2c}}}\int_{-{l}/{2}}^{{l}/{2}}{d{z}'}\hat{D}_{\sigma \pm }\left( {z}',\omega  \right){{e}^{\pm {\kappa}\omega {{z}'}/{c}}},
\end{equation}
where the Langevin noise operators ${{\hat{D}}_{\sigma \pm }}\left( z,\omega  \right)$ are previously defined in Eqs.~(\ref{the relations of operators Dx}) and~(\ref{the relations of operators Dy}).
In the second step, we relate the operators $\hat{a}_{\sigma \pm }^{(j+1)}\left( {{z}_{j}},\omega  \right)$ and $\hat{a}_{\sigma \pm}^{(j)}\left( {{z}_{j}},\omega  \right)$ in neighboring layers across the interface to each other by imposing the boundary conditions at $z = z_j \,\,(j=1,2)$, in which the tangential components of the electric and magnetic fields operators must be continuous (see Appendix~\ref{App:Boundary Conditions}). Therefore, after some manipulations, we arrive at:
\begin{equation}\label{operator relation in neighboring layers}
\left( \begin{matrix}
   \hat{a}_{\sigma +}^{\left( j+1 \right)}\left( {{z}_{j}},\omega  \right)  \\
   \hat{a}_{\sigma -}^{\left( j+1 \right)}\left( {{z}_{j}},\omega  \right)  \\
\end{matrix} \right)={\mathbb S}_{\sigma }^{\left( j \right)}\left( \begin{matrix}
   \hat{a}_{\sigma +}^{\left( j \right)}\left( {{z}_{j}},\omega  \right)  \\
   \hat{a}_{\sigma -}^{\left( j \right)}\left( {{z}_{j}},\omega  \right)  \\
\end{matrix} \right).
\end{equation}
Here, $z_{1(2)}=-(+) l/2$ and the elements of the transformation matrix
${\mathbb S}^j_\sigma$ are given in Appendix~\ref{App:Components of Transformation Matrix}. In the last step, by applying Eq.~(\ref{operators within slab}) and twice Eq.~(\ref{operator relation in neighboring layers}), we get the input-output relation for amplitude operators as follows:
\begin{eqnarray}\label{input - output quantum relation}
\left( \begin{matrix}
   \hat{a}_{\sigma -}^{\left( 1 \right)}\left( {-l}/{2},\omega  \right)  \\
   \hat{a}_{\sigma +}^{\left( 3 \right)}\left( {l}/{2},\omega  \right)  \\
\end{matrix} \right)&=&\left( \begin{matrix}
   {{R}_{\sigma }} & {{T}_{\sigma }}  \\
   {{T}_{\sigma }} & {{R}_{\sigma }}  \\
\end{matrix} \right)\left( \begin{matrix}
   \hat{a}_{\sigma +}^{\left( 1 \right)}\left( {-l}/{2},\omega  \right)  \\
   \hat{a}_{\sigma -}^{\left( 3 \right)}\left( {l}/{2},\omega  \right)  \\
\end{matrix} \right)
\nonumber\\
&&+ {\mathbb A}_{\sigma }\left( \begin{matrix}
   \hat{g}_{\sigma +}\left( \omega  \right)  \\
   \hat{g}_{\sigma -} \left( \omega  \right)  \\
\end{matrix} \right),
\end{eqnarray}
where the reflection and transmission
coefficients of the moving MDS,~${{R}_{\sigma }}$ and~${{T}_{\sigma }}$, are given by the classical expressions:
\begin{subequations}\label{reflection and transmission RT}
\begin{eqnarray}
{{R}_{\sigma }}&=&\frac{\left( {{e}^{2i{{n}_{\rm eff}}\left( \omega  \right)\,\omega {l}/{c}}}-1 \right)\left( n_{\rm eff}^{2}\left( \omega  \right)-\zeta_{\sigma }^{2}
\right){{e}^{-i\omega {l}/{c}}}}{{{\left( {\zeta_{\sigma }}+{{n}_{\rm eff}}\left( \omega  \right) \right)}^{2}}-{{\left( {\zeta_{\sigma }}-{{n}_{\rm eff}}\left( \omega
\right) \right)}^{2}}{{e}^{2i{{n}_{\rm eff}}\left( \omega  \right)\omega {l}/{c}}}},\,\,\,\,\,\,\,\,\,\,\,\,\,\label{the effects of the reflection }
\\
{{T}_{\sigma }}&=&\frac{4{{n}_{\rm eff}}\left( \omega  \right){\zeta_{\sigma }}{{e}^{-i\omega {l}/{c}}}{{e}^{i{{n}_{\rm eff}}\left( \omega  \right)\omega
{l}/{c}}}}{{{\left( {\zeta_{\sigma }}+{{n}_{\rm eff}}\left( \omega  \right) \right)}^{2}}-{{\left( {\zeta_{\sigma }}-{{n}_{\rm eff}}\left( \omega  \right)
\right)}^{2}}{{e}^{2i{{n}_{\rm eff}}\left( \omega  \right)\omega {l}/{c}}}}.\label{the effects of the transmission }
\end{eqnarray}
\end{subequations}
Here, the parameter~${\zeta_{\sigma }}\,\, (\sigma=x,y)$ is equal to ${{\mu }_{\rm eff ,yy}}$ and ${{\mu }_{\rm eff ,xx}}$ for the polarizations along $x$ and $y$ directions, respectively. In Eq.~(\ref{input - output quantum relation}), ${\mathbb A}_{\sigma }$ is the absorption matrix which arise from dissipative nature of the slab. The elements of this
$2\times 2$ matrix are given in Appendix~\ref{App:Components of Absorbing Matrix}.
Also, the quantum noise operators $\hat{g}_{\sigma \pm }(\omega)$ in Eq.~(\ref{input - output quantum relation}) are given by:
\begin{equation}\label{quantum noise operators of medium}
\hat{g}_{\sigma \pm }\left( \omega  \right)={{\left[ 2{{c}_{\sigma \pm }}\left( l,\omega  \right) \right]}^{-{1}/{2}}}\left[ {{{{\hat{g}}'}}_{\sigma
-}}\left( \omega  \right)\pm {{{{\hat{g}}'}}_{\sigma +}}\left( \omega  \right) \right],
\end{equation}
where
\begin{subequations}
\begin{eqnarray}
{g'_{x\pm }}\left( \omega  \right)&=&i\sqrt{\frac{\omega }{c}}{{e}^{i{{n}_{\rm eff}}\left( \omega  \right)\omega
{l}/{2c}}}\int_{-{l}/{2}}^{{l}/{2}}{d\,{z}'}\,{{e}^{\mp i{{n}_{\rm eff}}\left( \omega  \right){z}'{\omega }/{c}}}\nonumber\\
&&\times \Bigg[\frac{\sqrt{\varepsilon_{\rm
eff\,xx}^{I}}{{{\hat{f}}}_{e,x}}\left( {z}',\omega  \right)}{\sqrt{\varepsilon_{\rm eff ,xx}^{I}+E_m{{\left| {{n}_{\rm eff}}\left(
\omega  \right) \right|}^{2}}}} \nonumber \\
&& \,\,\,\,\,\pm \frac{{{n}_{\rm eff}}\left( \omega  \right)  \sqrt{E_m} {{\hat{f}}_{m,\bot }}\left( \pm z',\omega  \right)}{\sqrt{\varepsilon _{\rm eff ,xx}^{I}+E_m{{\left| {{n}_{\rm eff}}\left( \omega  \right) \right|}^{2}}}}\Bigg],\label{quantum noise operators 1}
\\
{g'_{y\pm }}\left( \omega  \right)&=&i\sqrt{\frac{\omega }{c}}{{e}^{i{{n}_{\rm eff}}\left( \omega  \right)\omega
{l}/{2c}}}\int_{-{l}/{2}}^{{l}/{2}}{d\,{z}'}\,{{e}^{\mp i{{n}_{\rm eff}}\left( \omega  \right){z}'{\omega }/{c}}}\nonumber\\
&&\times \Bigg[\frac{\sqrt{E_e}{{{\hat{f}}}_{e,\bot}}\left( {z}',\omega  \right)}{\sqrt{E_e -\mu_{\rm eff ,xx}^{-1\,I}{{\left| {{n}_{\rm eff}}\left(
\omega  \right) \right|}^{2}}}} \nonumber \\
&& \,\,\,\,\, \pm \frac{i{{n}_{\rm eff}}\left( \omega  \right)\sqrt{-\mu_{\rm eff ,xx}^{-1\,I}}{{{\hat{f}}}_{m ,x}}\left( {z}',\omega
\right)}{\sqrt{E_e -\mu_{\rm eff ,xx}^{-1\,I}{{\left| {{n}_{\rm eff}}\left( \omega  \right) \right|}^{2}}}}\Bigg],\label{quantum noise operators 1}
\end{eqnarray}
\end{subequations}
and
\begin{subequations}\label{coefficients cxcy}
\begin{eqnarray}
{{c}_{x\pm }}\left( l,\omega  \right)&=&{{e}^{-{\kappa}\omega {l}/{c}}}\Bigg(\frac{\sinh  \left( {{\kappa}\omega l}/{c} \right)}{{\kappa}}\\
&& \pm \frac{\varepsilon _{\rm eff ,xx}^{I}- E_m{{\left| {{n}_{\rm eff}}\left( \omega  \right) \right|}^{2}}}{\varepsilon _{\rm
eff ,xx}^{I}+E_m{{\left| {{n}_{\rm eff}}\left( \omega  \right) \right|}^{2}}}\frac{\sin \left( {\eta}{\omega }l/{c} \right)}{{\eta}}\Bigg),\nonumber\label{coefficients cx}
\\
{{c}_{y\pm }}\left( l,\omega  \right)&=&{{e}^{-{\kappa}\omega {l}/{c}}}\Bigg(\frac{\sinh \left( {{\kappa}\omega l}/{c} \right)}{{\kappa}}\\
&&\pm
\frac{{E_e }+\mu_{\rm eff ,xx}^{-1\,I}{{\left| {{n}_{\rm eff}}\left( \omega  \right) \right|}^{2}}}{{E_e }-\mu_{\rm eff ,xx}^{-1\,I}{{\left| {{n}_{\rm eff}}\left(
\omega  \right) \right|}^{2}}}\frac{\sin \left( {\eta}{\omega }l/{c} \right)}{\eta}\Bigg).\nonumber\label{coefficients cy}
\end{eqnarray}
\end{subequations}
By applying Eqs.~(\ref{commutation relations of new bosonic operator}) and making use of Eqs.~(\ref{quantum noise operators of medium})-(\ref{coefficients cxcy}), the quantum noise operators $\hat{g}_{\sigma \pm }\left( \omega  \right)$ are found to satisfy the following commutation relations:
\begin{subequations}\label{commutation relations of quantum noise operators }
\begin{eqnarray}
\left[ \hat{g}_{\sigma \pm } \left( \omega  \right),\hat{g}_{{\sigma }'\pm }^{\dagger }\left( {{\omega }'} \right) \right]&=&{{\delta }_{\sigma {\sigma }'}}\delta \left( \omega -{\omega
}' \right),
\\
\left[ \hat{g}_{\sigma \pm } \left( \omega  \right),\hat{g}_{{\sigma }'\mp }^{\dagger }\left( {{\omega }'} \right) \right]&=&0.
\end{eqnarray}
\end{subequations}
The equations~(\ref{input - output quantum relation})-(\ref{commutation relations of quantum noise operators }) make it possible to calculate the quantum statistical properties of the output fields at any position outside the slab, from
the properties of the input fields and the noise operators. Interestingly, the input-output relations~(\ref{input - output quantum relation}) are completely consistent with those of reported in~\cite{Dong2011,Hoseinzadeh2017} for the input-output relation of a stationary anisotropic slab with ${{\varepsilon }_{yz}}={{\mu }_{yz}}=0$. Because, as stated earlier, an isotropic moving MDS behaves somewhat like an anisotropic MDS.

To compare the input-output relations of the moving MDS~(\ref{input - output quantum relation}) with the corresponding relations for a dielectric slab at rest~\cite{Gruner1996,Artoni1998a,Matloob2000,Amooghorban2013,Amooghorban arXiv,Hoseinzadeh2017}, we define new noise operators as: $\left( \begin{matrix}
   \hat{F}_{\sigma +}\left( \omega  \right)  \\
   \hat{F}_{\sigma -}\left( \omega  \right)  \\
\end{matrix} \right)$=
$A_{\sigma }\left( \begin{matrix}
   \hat{g}_{\sigma +}\left( \omega  \right)  \\
   \hat{g}_{\sigma -}\left( \omega  \right)  \\
\end{matrix} \right)$.
They represent the quantum noises associated with
loss inside the moving MDS, and have the following expectation values at the finite temperature $\Theta$:
\begin{equation*}\label{New noise operators expectation value}
\langle \left. F \right|\hat{F}_{\sigma \pm }^{\dagger }\left( \omega  \right)\left| F\rangle =\langle \left. F \right| \right.{{\hat{F}}_{\sigma \pm }}\left|
F\rangle =0 \right.,
\end{equation*}
\begin{eqnarray}\label{New noise operators expectation value}
\langle \left. F \right|\hat{F}_{\sigma \pm }^{\dagger }\left( \omega  \right){{\hat{F}}_{\sigma \pm }}\left( {{\omega }'} \right)\left| F\rangle \right.& =& N( \gamma\omega ,\Theta )
\\
& \times &(1-{{\left| {{R}_{\sigma }} \right|}^{2}}-{{\left| {{T}_{\sigma }} \right|}^{2}})\delta \left( \omega -{\omega }' \right),\nonumber
\end{eqnarray}
where $|F \rangle$ represents the noise state of the MDS, and $N\left( \omega ,\Theta \right)={{[\exp ({\hbar \omega }/{{{k}_{B}}\Theta}\;)-1]}^{-1}}$ is the mean number of thermal photons at frequency $\omega$ and temperature $\Theta$, in which $\hbar $ is the Planck constant per $2\pi$ and ${{k}_{B}}$ is the Boltzmann's constant.
Since the input fields in the free space cannot sense the presence of the moving MDS before arrive at it, the optical input operators must satisfy the usual bosonic commutation relations as:
\begin{eqnarray}\label{commutation relations of input operators}
\left[ \hat{a}_{\sigma +}^{(1)}\left( z,\omega  \right),\hat{a}_{{\sigma }'+}^{(1)\dagger }\left( {z}',{\omega }' \right) \right]&=&\left[ \hat{a}_{\sigma
-}^{(3)}\left( z,\omega  \right),\hat{a}_{{\sigma }'-}^{(3)\dagger }\left( {z}',{\omega }' \right) \right]\nonumber \\
&=&{{\delta }_{\sigma {\sigma }'}} \delta \left( \omega -{\omega }' \right) .\,\,\,\,\,\,
\end{eqnarray}
By using the above commutation relations and the input-output relations~(\ref{input - output quantum relation}), the bosonic commutation relations
for the outgoing operators are obtained as follows:
\begin{subequations}
\begin{eqnarray}
\left[ \hat{a}_{\sigma -}^{(1)}\left( z,\omega  \right),\hat{a}_{{\sigma }'-}^{(1)\dagger }\left( {z}',{\omega }' \right) \right]&=&\left[ \hat{a}_{\sigma
+}^{(3)}\left( z,\omega  \right),\hat{a}_{{\sigma }'+}^{(3)\dagger }\left( {z}',{\omega }' \right) \right]\nonumber\\
&=&{{\delta }_{\sigma {\sigma }'}} \delta ( \omega -{\omega }' ),\,\,\,\,\,\,\,\,\,\,\,\,\,\label{commutation relations of output modes operators 1}
\\
\left[ \hat{a}_{\sigma -}^{(1)}\left( z,\omega  \right),\hat{a}_{{\sigma }'+}^{(3)\dagger }\left( {z}',{\omega }' \right) \right]&=&\left[ \hat{a}_{\sigma
+}^{(3)}\left( z,\omega  \right),\hat{a}_{{\sigma }'-}^{(1)\dagger }\left( {z}',{\omega }' \right) \right]\nonumber \\
&=&0.\label{commutation relations of output modes operators 2}
\end{eqnarray}
\end{subequations}

In the limiting case that the moving MDS is at rest, that is $v= 0$, the effective tensors ${{{\bar{\bar{\boldsymbol\varepsilon }}}}_{\rm eff}}$ and ${{{\bar{\bar{\boldsymbol\mu }}}}_{\rm eff}}$ reduce to the stationary parameters $\varepsilon$ and $\mu$. Consequently, the input-output relations~(\ref{input - output quantum relation}) for $\mu=1$ tend to those derived in~\cite{Artoni1998a,Artoni1998b,Matloob2000} for a stationary dielectric slab.
%
\section{Numerical Calculations and Analysis}\label{Numerical Calculation and Analysis}

%
\begin{figure*}[t]
\includegraphics[width = 0.64\columnwidth]{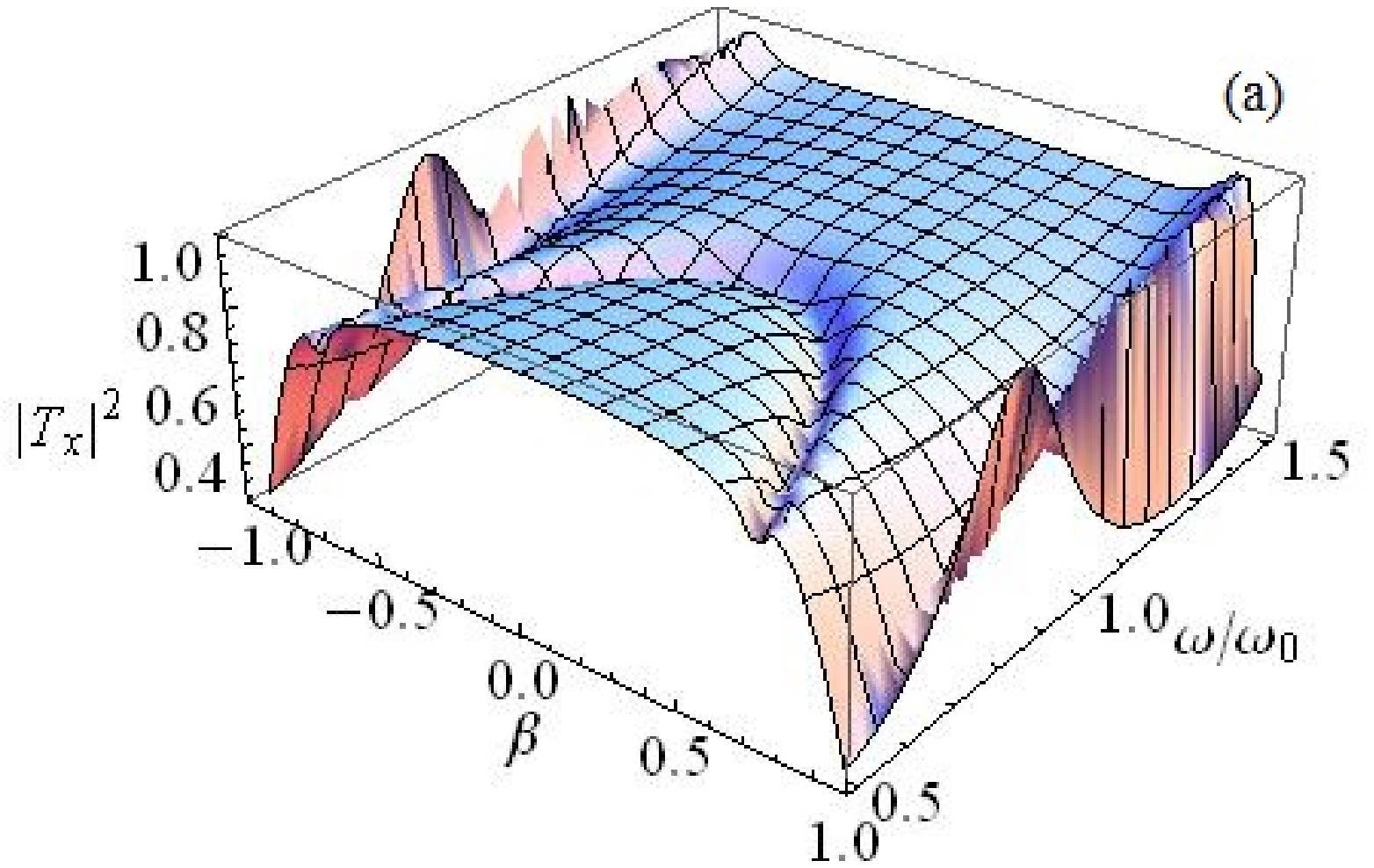}
\includegraphics[width = 0.64\columnwidth]{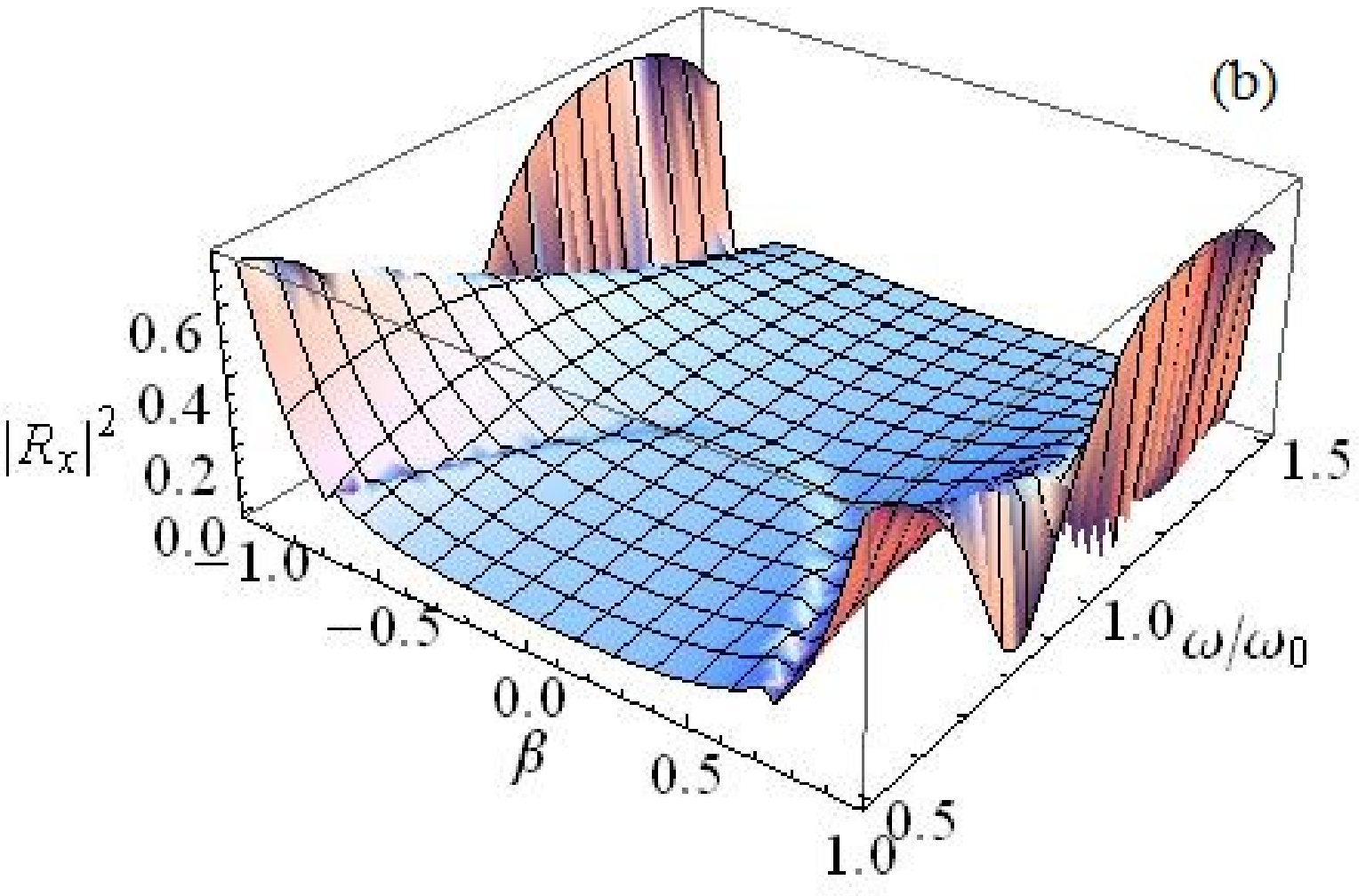}
\includegraphics[width = 0.76\columnwidth]{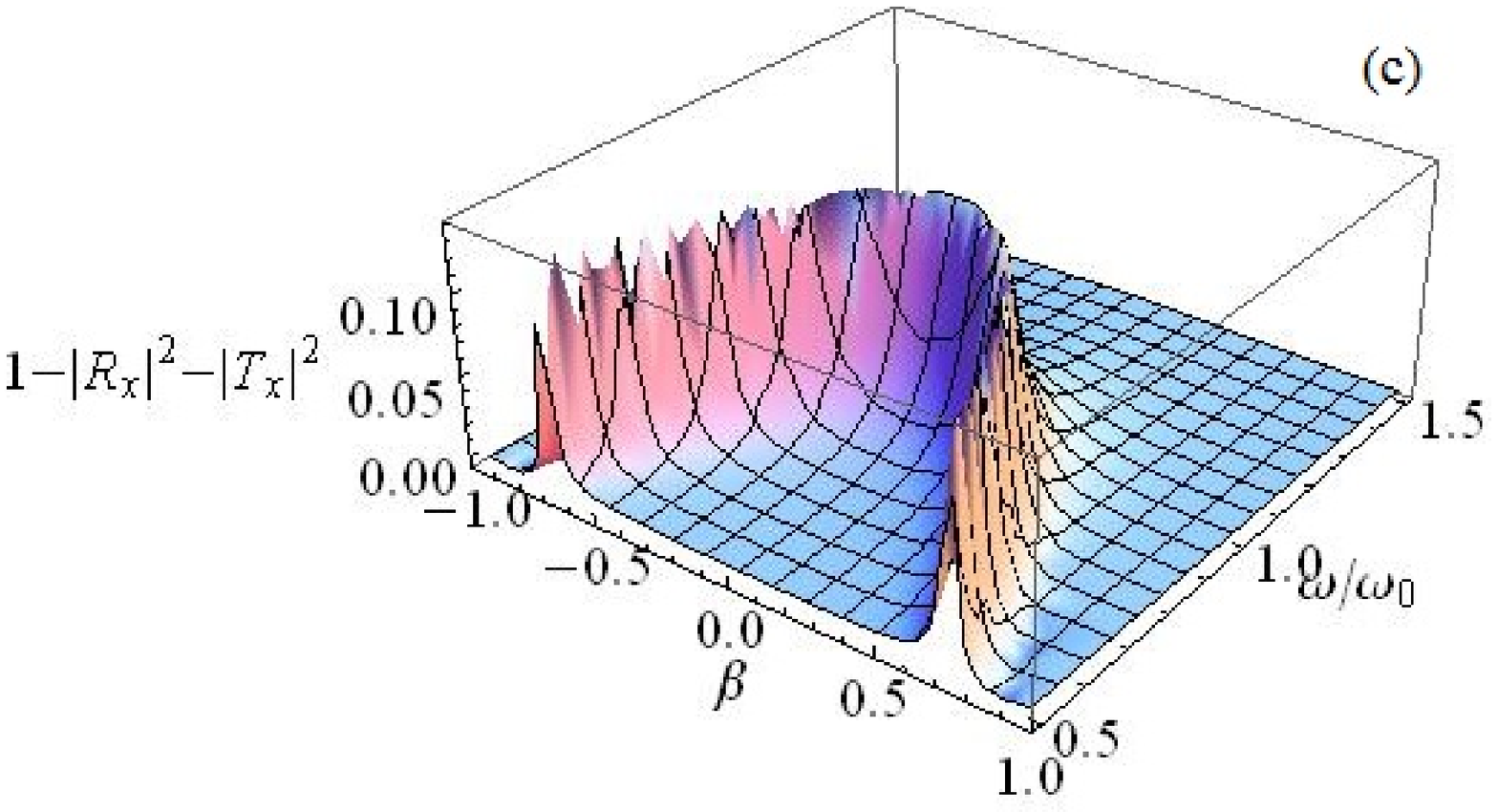}
\caption{ (a) The square modulus of the transmission coefficient $|T_x|^2$, (b) the reflection coefficient $|R_x|^2$, and (c) the absorption coefficient $1-|R_x|^2-|T_x|^2$ for x-polarized light incident on the moving MDS with thickness ${{{\omega}_{0}}l}/{c}=1$. The permittivity and permeability functions of the MGS in its rest frame are described by the Lorentz model~(\ref{Lorentz model}), with parameters: $\varepsilon_\infty=2$, $\mu_\infty=1$, ${{{\gamma
}_{e}}}/{{{\omega }_{0}}}\;=0.1$, ${{{\gamma
}_{m}}}/{{{\omega }_{0}}}\;=0.2$, ${{{\omega }_{p e}}}/{{{\omega }_{0}}}\;=0.1$,  and ${{{\omega }_{p m}}}/{{{\omega }_{0}}}\;=0.05$.}
\label{Fig:transmission, reflection and absorption coefficients for x-polarized incidence}
\end{figure*}
\begin{figure*}[t]
\includegraphics[width =0.64\columnwidth]{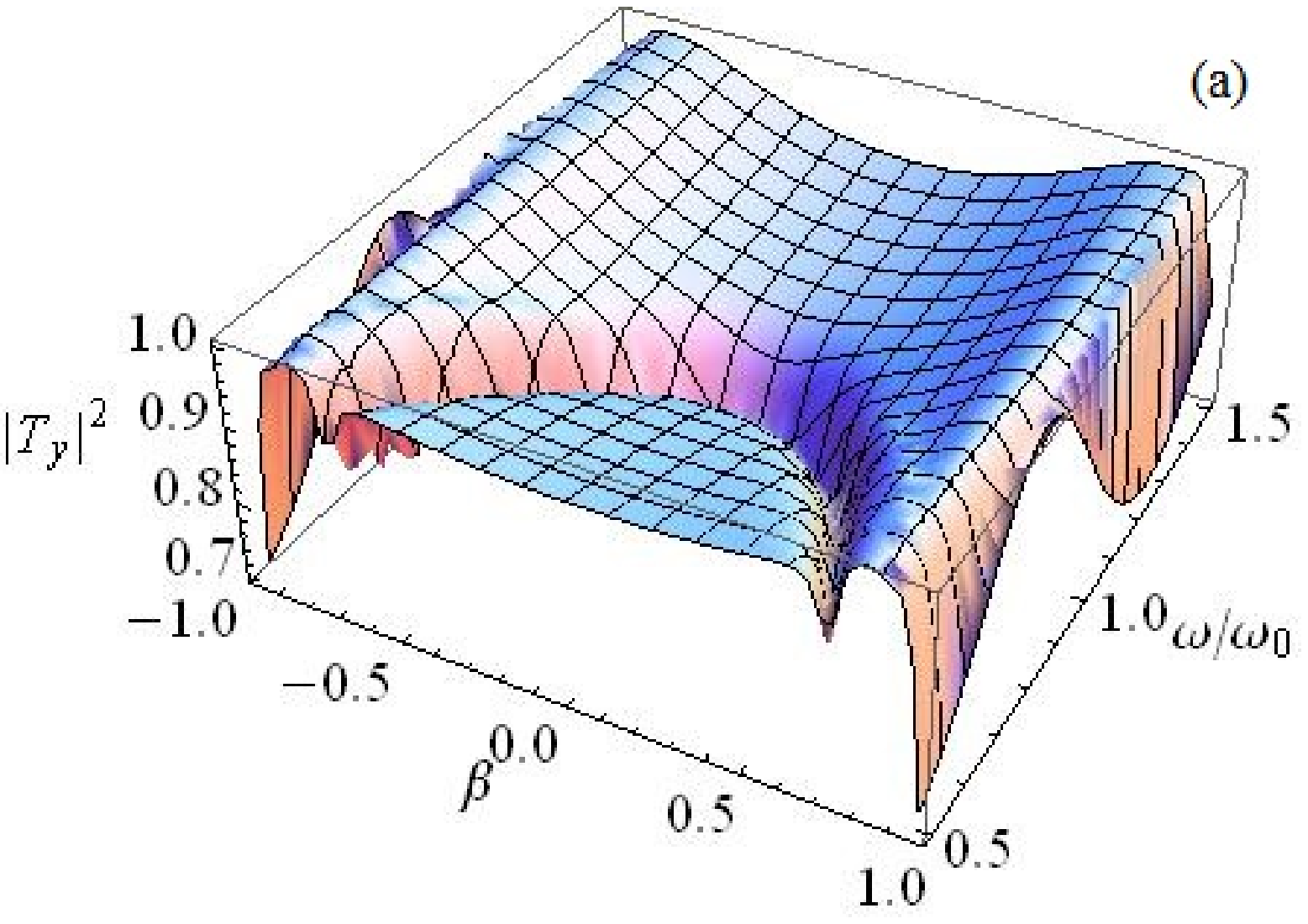}
\includegraphics[width =0.64\columnwidth]{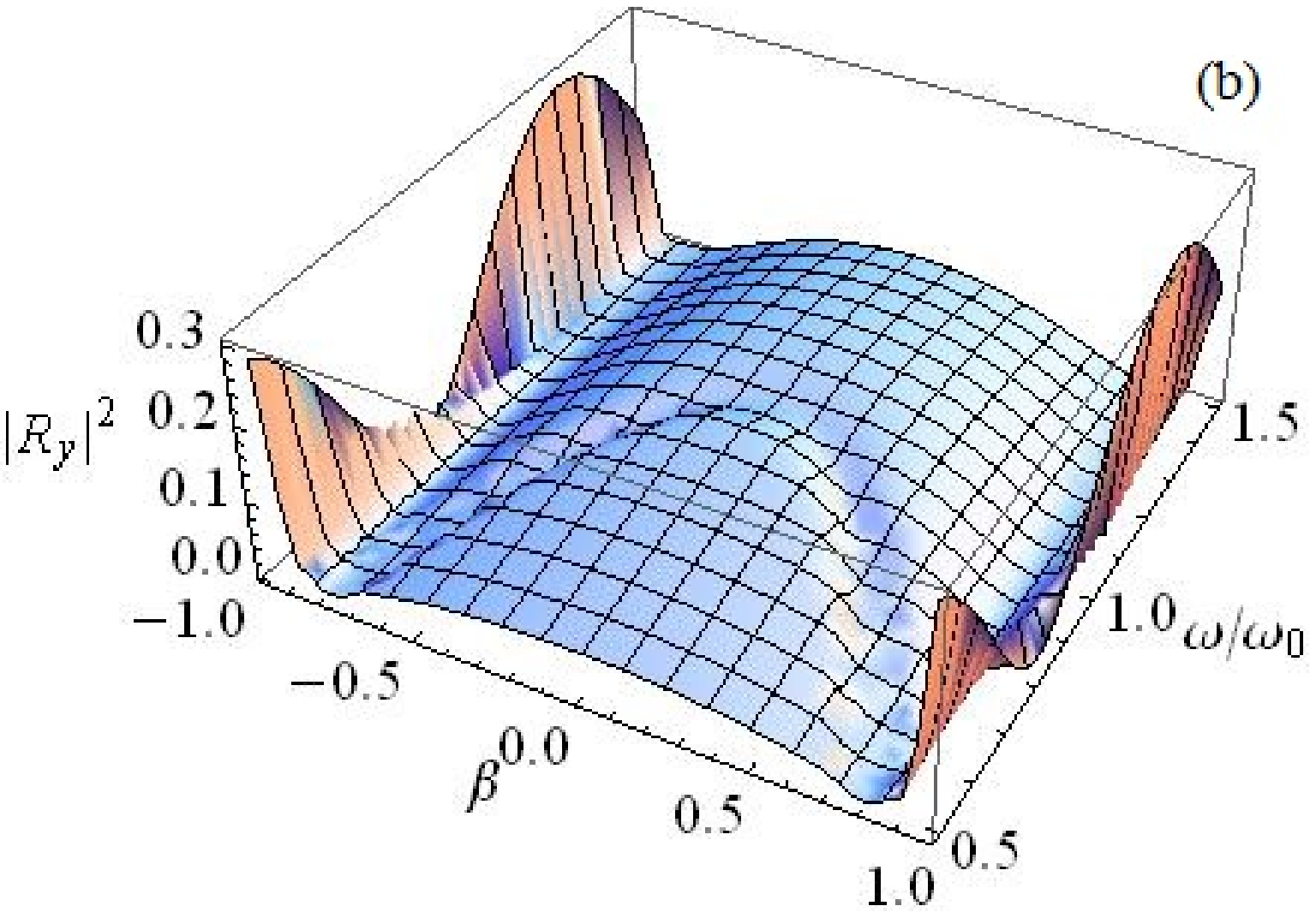}
\includegraphics[width = 0.76\columnwidth]{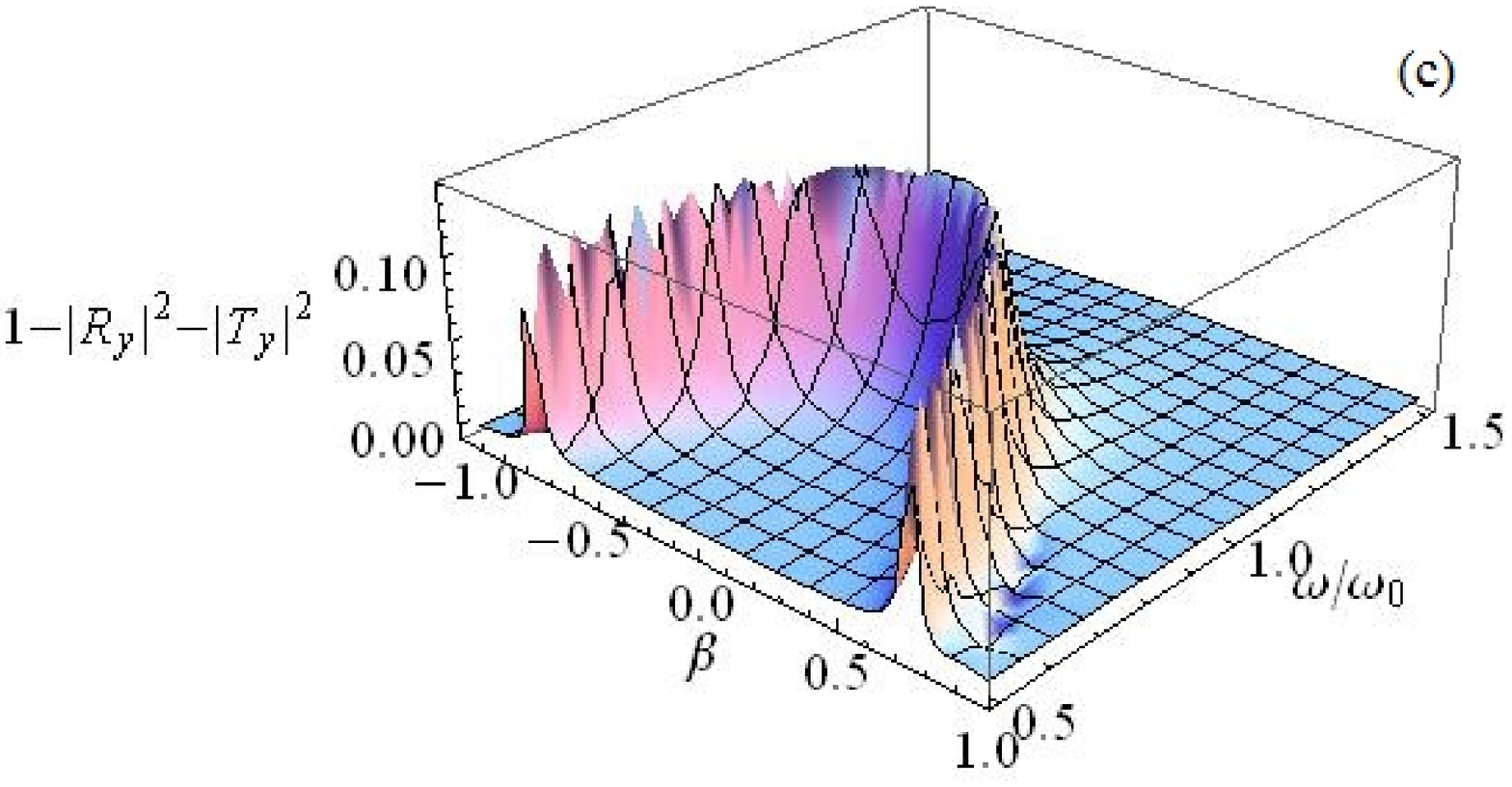}
\caption{Same as Fig.~\ref{Fig:transmission, reflection and absorption coefficients for x-polarized incidence} but now for $y$-polarized light incident on the moving MDS.}
\label{Fig:transmission, reflection and absorption coefficients for y-polarized incidence}
\end{figure*}

\subsection{Transmission, reflection and absorption coefficients}
%
Due to the complexity of Eq.~(\ref{reflection and transmission RT}), it is difficult to predict the results analytically. We start our numerical analysis by examining the motion effects of the moving MDS on the transmission and reflection properties. Consider that a single-resonance MDS of Lorentz type whose complex permittivity and permeability functions in the
rest frame of the MDS are given by~\cite{Amooghorban2013,Amooghorban arXiv,Amooghorban2014}
\begin{subequations}\label{Lorentz model}
\begin{eqnarray}
\varepsilon(\omega )= \varepsilon_\infty \Big(1+\frac{\omega_{pe}^2}{\omega_{0}^{2}-{\omega }^{2}-i\gamma_e \omega }\Big),\label{Lorentz permittivity}
\\
\mu(\omega )= \mu_\infty \Big(1+\frac{\omega_{pm}^2}{\omega_{0}^{2}-{\omega }^{2}-i\gamma_m \omega }\Big),\label{Lorentz permeability}
\end{eqnarray}
\end{subequations}
where $\varepsilon_\infty$ and $\mu_\infty$ are, respectively, the high-frequency limit of $\varepsilon$ and $\mu$, ${{\omega }_{p}}$ and ${{\omega }_{0}}$ are, respectively, the plasma and the resonance frequency, and $\gamma $ is the absorbtion coefficient of the MDS.

In Figs.~\ref{Fig:transmission, reflection and absorption coefficients for x-polarized incidence} and~\ref{Fig:transmission, reflection and absorption coefficients for y-polarized incidence}, the square modulus of the transmission and reflection coefficients, $|T_x|^2$ and $|R_x|^2$, and the absorption coefficient, $1-|R_x|^2-|T_x|^2$, for x-polarized light are plotted as functions of dimensionless parameters ${\omega }/{{{\omega }_{0}}}\;$ and $\beta $. Here, positive(negative) value of $\beta$ represents that the MDS is moving in the positive(negative) y-direction.
It is seen that all plots are symmetric with respect to $\beta=0$, because these optical coefficients depend on the even powers of $\beta$.
Furthermore, as $\beta$ varies from $0$ to $ 1$, $|R_x|^2$ increases very slowly with $\beta$ and then shows an oscillatory behavior followed by a fast enhancement to $1$, in the limit of $\beta \rightarrow 1$. On the contrary, $|T_x|^2$ decreases very slowly and then shows an oscillatory behavior followed by a fast decrease to zero, as $\beta$ approaches to $1$. Similar behaviors are seen in Figs.~\ref{Fig:transmission, reflection and absorption coefficients for y-polarized incidence}(a) and~\ref{Fig:transmission, reflection and absorption coefficients for y-polarized incidence}(b) for the y-polarized light incident on the moving MDS .

In the low velocity limit $v\ll c\,\,(\beta \ll 1)$, the absorption coefficient reaches the maximum value of $0.14$ near the resonance frequency. While, this maximum value shifts to frequencies below the resonant frequency of the MDS with increasing $\beta$, as seen in Figs.~\ref{Fig:transmission, reflection and absorption coefficients for x-polarized incidence}(c) and~\ref{Fig:transmission, reflection and absorption coefficients for y-polarized incidence}(c).

In the ultra-relativistic limit $v\simeq c\,\, (\beta\simeq1)$, the relation $|R_x|^2+|T_x|^2\approx 1$ holds, and the moving MDS acts like a lossless slab,
with $|T_x|^2\approx 0$ and $|R_x|^2\approx 1$, i.e., the moving MDS behaviors as a perfectly conducting slab to the incident quantum light. This is in confirmation with the results obtained in the work of~\cite{Huang1994}.
%

\subsection{Quadrature squeezing}\label{SubSec:quadrature squeezing}
%
\begin{figure*}[t]
\includegraphics[width = 0.9\columnwidth]{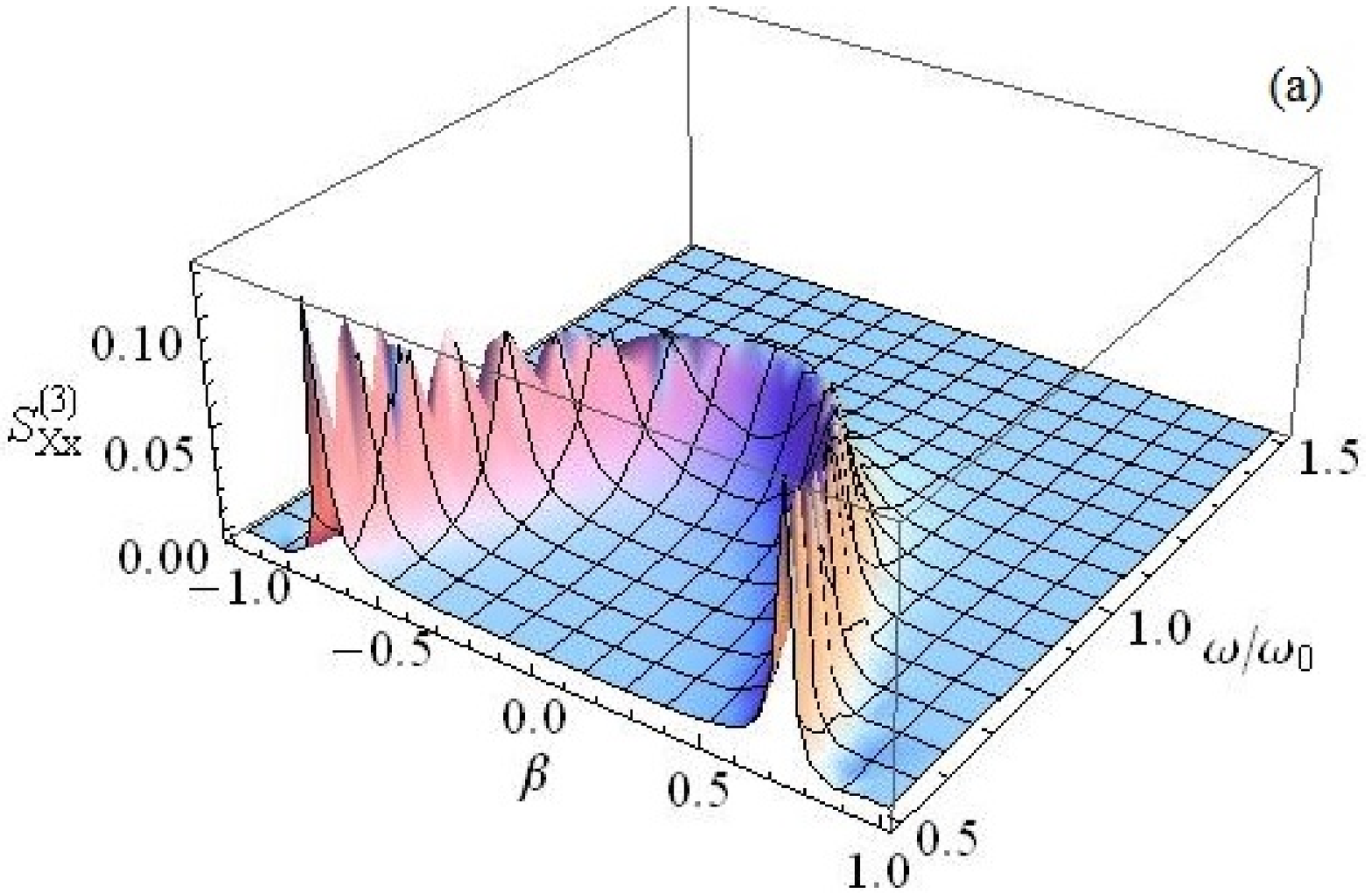}
\includegraphics[width = 0.9\columnwidth]{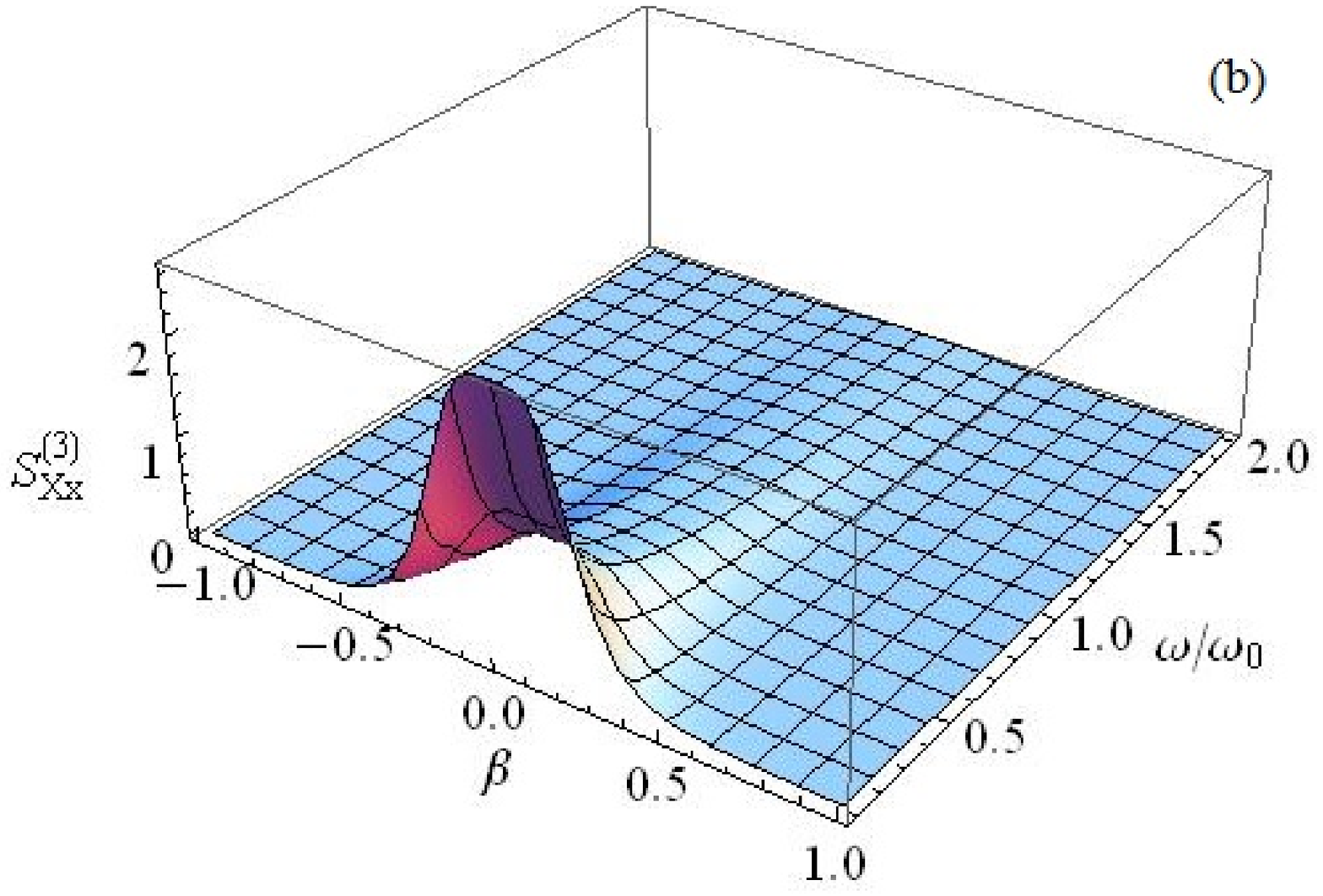}
\caption{ (a) The squeezing parameter $S_{Xx}^{(3)}$ as functions of dimensionless parameters  ${\omega }/{{{\omega }_{0}}}$ and $\beta$ for the transmitted CS through the moving MDS at temperature of $\hbar\omega_0/k_B\Theta=10/6$. (b) The squeezing variance $S_{Xx}^{(3)}$ as function of dimensionless parameters  $\hbar\omega_0/k_B\Theta$ and $\beta$ for the transmitted CS through the moving MDS at fixed frequency of ${\omega }/{{{\omega }_{0}}}=1$. The material properties of the moving MDS slab in its rest frame are described by Lorentz model~(\ref{Lorentz model}) with parameters are identical to those used in Fig.~\ref{Fig:transmission, reflection and absorption coefficients for x-polarized incidence}. Here, the mean number of photons in the coherent state $|\alpha_x\rangle_R$ is 16.}
\label{Fig:squeezing parameter}
\end{figure*}

In this subsection, we shall proceed to study the significant impacts of the motion of the moving MDS on the noise properties
of the transmitted states.
To do so, let us start with the definition of the quadrature operators of the output field in the region $z>l/2$ as follows:
\begin{subequations}
\begin{eqnarray}
{{\hat{X}}^{(3)}_{\sigma}}=\frac{1}{2}\left( \hat{a}_{\sigma +}^{\left( 3 \right)}+\hat{a}_{\sigma +}^{\left( 3 \right)\dagger } \right),\label{the quadrature uncertainty x}
\\
{{\hat{Y}}^{(3)}_{\sigma}}=\frac{i}{2}\left( \hat{a}_{\sigma +}^{\left( 3 \right)\dagger }-\hat{a}_{\sigma +}^{\left( 3 \right)} \right).\label{the quadrature uncertainty y}
\end{eqnarray}
\end{subequations}
These quadratures are the analogues of the dimensionless position and momentum operators and are subject to a similar uncertainty relation $\langle \big(\Delta{{\hat{X}}_{\sigma}}^{(3)} \big)^2 \big(\Delta{{\hat{Y}}_{\sigma}}^{(3)} \big)^2\rangle >1/16$, where the variance of the arbitrary operator $\hat{\cal O}$ is defined as $\langle\Delta{{\hat{\cal O}}}^2\rangle=\langle{{\hat{\cal O}}}^2\rangle - \langle {{\hat{\cal O}}} \rangle^2$.
We can now quantify the quantum fluctuations of the transmitted light through the moving MDS by evaluating the variance of these field quadratures. Consider that the incident fields from the free space to the left and the right sides of the MDS are, respectively, the single mode CS, $|\alpha_\sigma\rangle_R$, and the conventional quantum vacuum state, $|0\rangle_L$, where the subscript indices $R$ and $L$ denote the direction of propagation. Therefore, the general state of the system is written as $\left| \psi\rangle  \right.=\left| {{\alpha }_{\sigma }} \right.{{\rangle }_{R}}\left| 0{{\rangle }_{L}} \right.\left| F\rangle  \right.$.
By using Eqs.~(\ref{input - output quantum relation}) and evaluating the variance of ${{\hat{X}}_{\sigma}}^{(3)}$ with respect to the state $|\psi\rangle$, after
some manipulations, we obtain:
\begin{subequations}\label{Delta xy}
\begin{eqnarray}
\langle \big({{\Delta \hat{X}_{\sigma}^{\left( 3 \right)} }}\big)^2\rangle=\frac{1}{4}\left( 1+2\langle \hat{F}_{\sigma +}^{\dagger }{{{\hat{F}}}_{\sigma +}}\rangle
\right),\label{Delta x}
\\
\langle \big({{\Delta \hat{Y}_{\sigma}^{\left( 3 \right)} }}\big)^2\rangle=\frac{1}{4}\left( 1+2\langle \hat{F}_{\sigma +}^{\dagger }{{{\hat{F}}}_{\sigma +}}\rangle
\right).\label{Delta y}
\end{eqnarray}
\end{subequations}
It is clearly seen that the transmitted CS through the moving MDS is not a CS due to the presence of the noise flux $\langle \hat{F}_{\sigma +}^{\dagger }{{{\hat{F}}}_{\sigma +}}\rangle$ at the right hand side of Eq.~(\ref{Delta xy}).
Of course, this noise flux vanishes at zero temperature, or at frequencies far from the resonant frequency of the MDS, or at extreme velocities where the moving MDS acts like a lossless slab. In these limiting cases, the output state stays in a minimum uncertainty state.

In what follows, for simplicity, we use the squeezing parameter, $S_{X\sigma}^{(3)}=4{{ \Delta \hat{X}_{\sigma}^{\left( 3 \right)2} }}-1$, to investigate the quadrature squeezing of the output light. This parameter is zero for the quantum vacuum and the coherent states and the existence of the quadrature squeezing (squeezing in the form of reduced quantum noises with respect to standard limit) is manifested in a negative-valued variance.

Fig.~\ref{Fig:squeezing parameter} shows the squeezing parameter $S_{X\sigma}^{(3)}$ for the transmitted CS with $ x$ polarization as functions of the dimensionless parameters $\beta $ and ${\omega }/{{{\omega }_{0}}}$. Here, we adopt the Lorentz model~(\ref{Lorentz model}) to characterize the dissipative and dispersive effects of the MDS in its rest frame.
Since, the fluctuations in other quadrature operator $S_{Y\sigma}^{(3)}$ follow a similar behavior, we confine our attention to the squeezing parameter $S_{X\sigma}^{(3)}$.

At the extreme velocity $v\simeq c$, as mentioned in the above, the moving MDS acts as a perfectly conducting slab to the input states. Therefore, it is expected that the output field is prepared in a state close to the quantum vacuum state, as seen in Fig.~\ref{Fig:squeezing parameter}(a).
On the contrary, at frequency ranges where the relation $|R_x|^2+|T_x|^2\approx 1$ holds except at extreme velocities, the transmitted field is prepared in a state close to a single mode CS. Therefore, in both above situations, the output state is prepared in a minimum uncertainty state, and consequently the squeezing parameter $S_{X\sigma}^{(3)}$ becomes zero.


To compare the thermal effects of the moving MDS with its motion effects, the squeezing parameter $S_{X\sigma}^{(3)}$ at the resonance frequency $\omega_0$ is shown in Fig.~\ref{Fig:squeezing parameter}(b) with respect to the dimensionless parameters $\beta $ and ${\hbar {{\omega }_{0}}}/{{{k}_{B}}T}$.
At elevated temperatures, we observe that the squeezing parameter reaches the maximum value around $\beta=0$, where the noise flux is very significant at the resonance frequency, and then decreases to zero with increasing $\beta$, because the absorption becomes very small, in the limit of $\beta \rightarrow 1$ (see Fig.~\ref{Fig:transmission, reflection and absorption coefficients for x-polarized incidence}(c)). Therefore, at low velocities and also at the resonance frequency, the thermal effects have the overall effect of degrading the quantum features of the input state.
%


\subsection{Mandel parameter }\label{SubSec:Mandel parameter}
\begin{figure*}[t]
\includegraphics[width = 0.9\columnwidth]{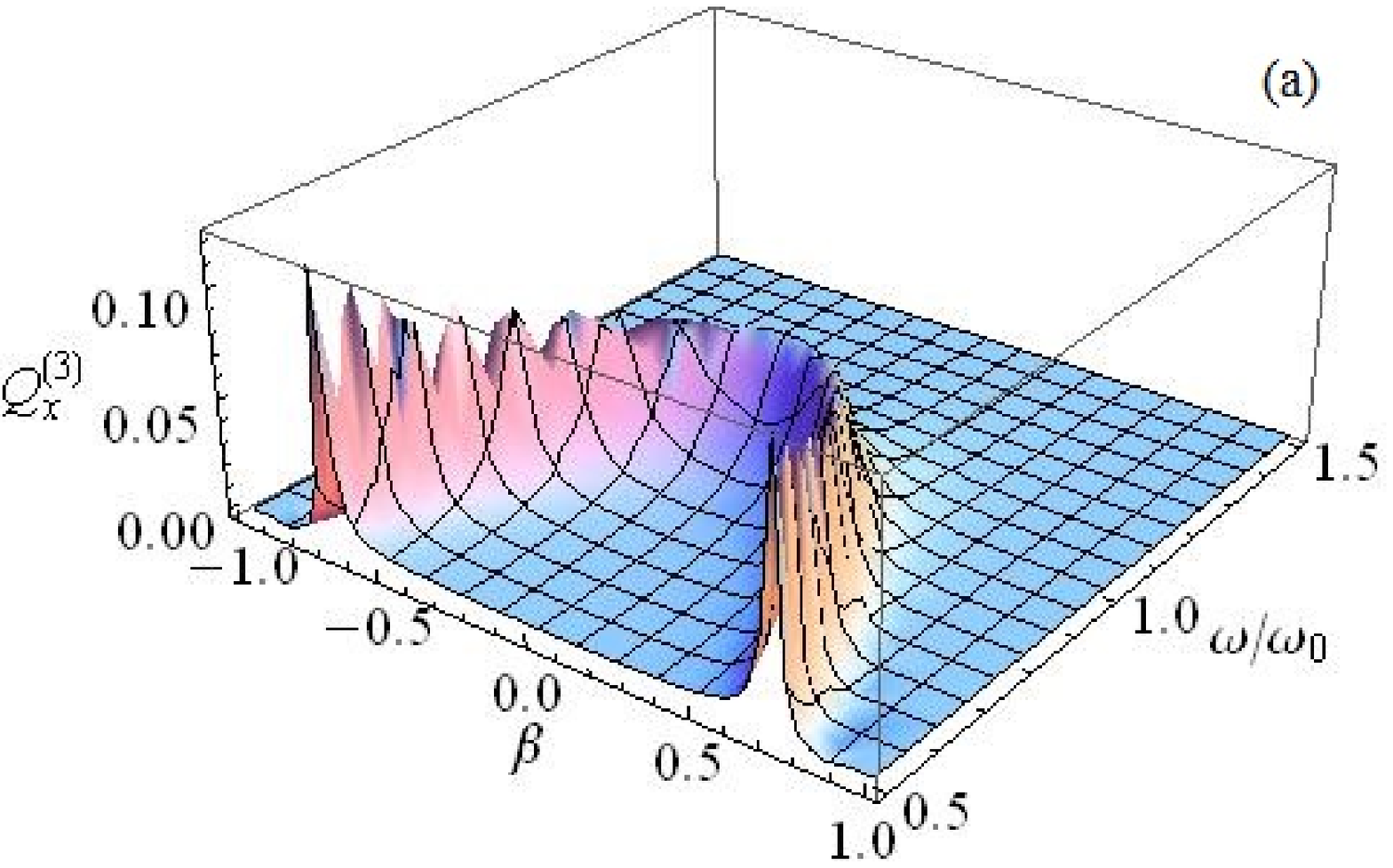}
\includegraphics[width = 0.9\columnwidth]{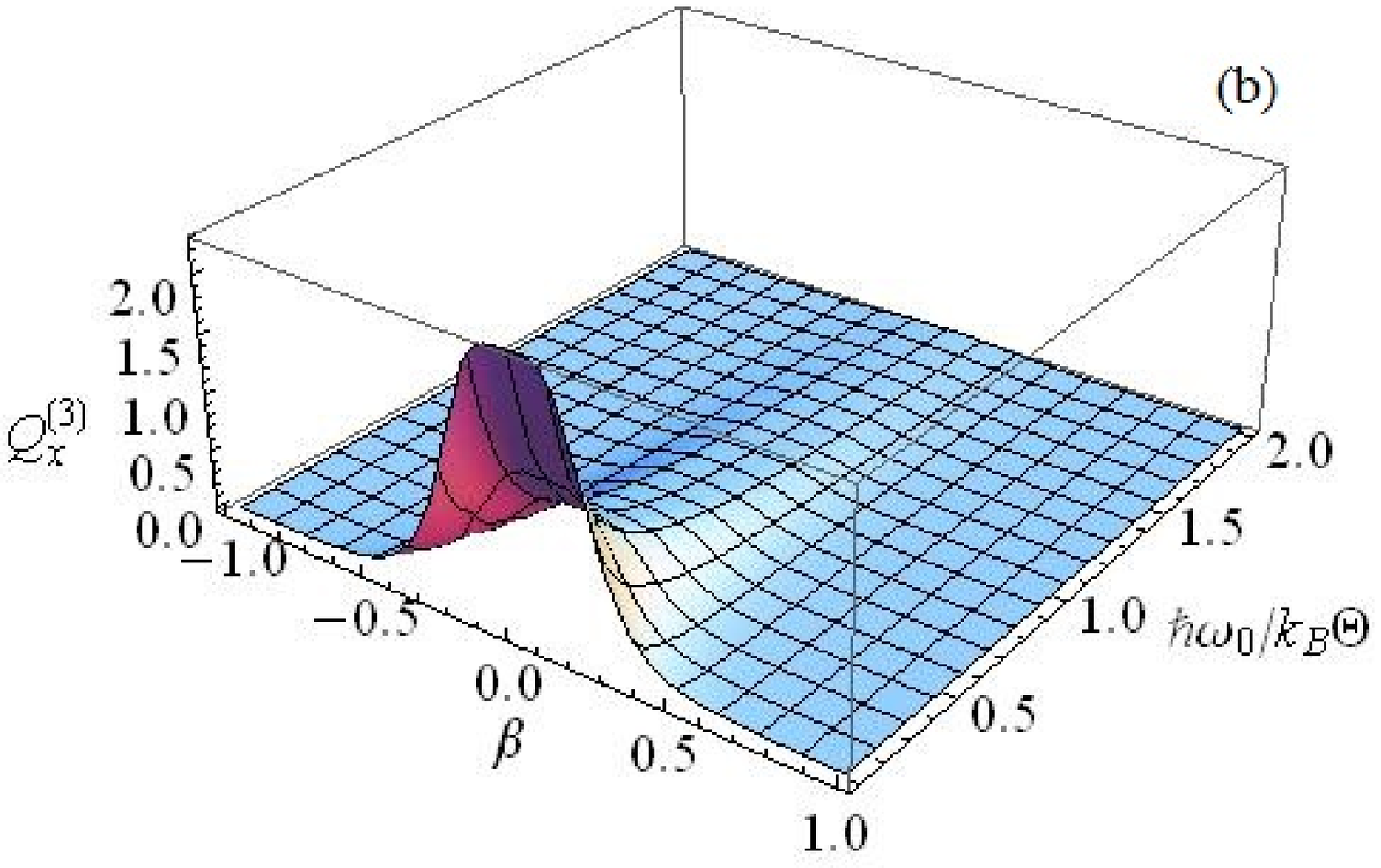}
\caption{Mandel parameter $Q_{x}^{(3)}$ as functions of dimensionless parameters  ${\omega }/{{{\omega }_{0}}}$ and $\beta$ for the transmitted CS through the moving MDS at temperature of $\hbar\omega_0/k_B\Theta=10/6$. (b) Mandel parameter $Q_{x}^{(3)}$ as function of dimensionless parameters  $\hbar\omega_0/k_B\Theta$ and $\beta$ for the transmitted CS through the moving MDS at fixed frequency of ${\omega }/{{{\omega }_{0}}}=1$. The material parameters are identical to those used in Fig.~\ref{Fig:transmission, reflection and absorption coefficients for x-polarized incidence}.}
\label{Fig:Mandel parameter}
\end{figure*}

In order to study the photon-counting statistics of the transmitted CSs through the MDS at the finite temperature,
we analyze the Mandel parameter~\cite{Scully1997}
\begin{equation}\label{Mandel parameter}
Q_{\sigma +}^{\left( 3 \right)}=\frac{\langle \hat{a}_{\sigma +}^{\left( 3 \right)\dagger }\hat{a}_{\sigma +}^{\left( 3 \right)}\hat{a}_{\sigma +}^{\left( 3
\right)\dagger }\hat{a}_{\sigma +}^{\left( 3 \right)}\rangle -{{\langle \hat{a}_{\sigma +}^{\left( 3 \right)\dagger }\hat{a}_{\sigma +}^{\left( 3 \right)}\rangle
}^{2}}-\langle \hat{a}_{\sigma +}^{\left( 3 \right)\dagger }\hat{a}_{\sigma +}^{\left( 3 \right)}\rangle }{\langle \hat{a}_{\sigma +}^{\left( 3 \right)\dagger
}\hat{a}_{\sigma +}^{\left( 3 \right)}\rangle },
\end{equation}
where the positive, zero and negative values of this parameter represents super-Poissonian, Poissonian and sub-Poissonian distribution, respectively~\cite{Scully1997}. By using Eq.~(\ref{input - output quantum relation}), after
some algebra, the Mandel parameter~(\ref{Mandel parameter}) for the output state
in the region $z>l/2$ can be obtained as:
\begin{eqnarray}\label{Q}
Q_{\sigma}^{\left( 3 \right)}&=&\frac{2{{\left| {{T}_{\sigma }}{{\alpha }_{\sigma }} \right|}^{2}}N\left( \gamma\omega ,\Theta \right)\left( 1-{{\left| {{T}_{\sigma }}
\right|}^{2}}-{{\left| {{R}_{\sigma }} \right|}^{2}} \right)}{{{\left| {{T}_{\sigma }}{{\alpha }_{\sigma }} \right|}^{2}}+N ( \gamma\omega ,\Theta )\left( 1-{{\left| {{T}_{\sigma
}} \right|}^{2}}-{{\left| {{R}_{\sigma }} \right|}^{2}} \right)}\nonumber \\
&&-\frac{\Big( N( \gamma\omega ,\Theta )\big({ 1-\left| T_\sigma \right|^2-\left|
R_\sigma \right|^2 }\big)\Big)^2}{{{\left| {{T}_{\sigma }}{{\alpha }_{\sigma }} \right|}^{2}}+N ( \gamma\omega ,\Theta )\left( 1-{{\left| {{T}_{\sigma
}} \right|}^{2}}-{{\left| {{R}_{\sigma }} \right|}^{2}} \right)},\,\,\,\,\,\,
\end{eqnarray}
With the help of the Lorentz model~(\ref{Lorentz model}), the motion effect of the moving MDS on the photon counting statistics of the transmitted x-polarized CS is shown in Fig.~\ref{Fig:Mandel parameter}.
%

In Fig.~\ref{Fig:Mandel parameter}(a), in confirmation of our recent findings in previous subsections, we observe that the Mandel parameter is positive only around regions where the absorption is significant, i.e., the transmitted CS exhibits super-Poisson distribution. Furthermore, at the frequencies where the relation $|R_x|^2+|T_x|^2\approx 1$ holds except at extreme velocities, the Mandel parameter becomes zero. Therefore, as far as it is related to the photon-counting statistics, the transmitted CS through the moving MDS is the same CS.

In Fig.~\ref{Fig:Mandel parameter}(b), we have plotted the Mandel parameter $Q_{x}^{\left(3 \right)}$ at the resonant frequency as functions of the dimensionless parameters $\beta $ and ${\hbar {{\omega }_{0}}}/{{{k}_{B}}T}$.
As it is seen, the thermal effects are predominant only in the low velocity $v<0.2c$, where the absorption is significant. In the high velocities, the thermal effects is minimal and subsequently the Mandel parameter becomes zero, because the moving MDS with high velocity behaves like a lossless slab at the resonant frequency [see Fig.~\ref{Fig:transmission, reflection and absorption coefficients for x-polarized incidence}(c)].

\section{Conclusions}\label{Sec:summary}

We have developed a phenomenological scheme for the quantization of the electromagnetic field propagating perpendicularly to a moving MDS with uniform velocity in the direction parallel to its interface. We have derived Input-output relations for this MDS and described the action of a moving lossy slab on an arbitrary quantum state of light with either $s$- or $p$-polarization. We have investigated the impact of the motion of the moving MDS on quantum
properties of the incident states. To this end, the input-output relations are used to investigate the impact of the motion of the moving MDS on the quantum
properties of the incident states.

By modeling the dispersive and dissipative effects of the moving MDS by the Lorentz model in its rest frame, we have evaluated
the quadrature squeezing and the Mandel parameter for the transmitted CS through the moving MDS. We found that the moving MDS at the extreme velocity acts as a perfectly conducting slab to the input states. Therefore, the output field is prepared in a state close to the quantum vacuum state. While, at the low and moderate velocities, $v<0.8c$, and also at the frequencies regions where the absorbtion is weak, the transmitted field is prepared in a state close to a single mode CS. It is also shown that the thermal effects have significant role in degrading the quantum features of the input state only at the low and moderate velocities.

\appendix
\include{app1}
%
\section{Boundary Conditions }\label{App:Square root of tensors}
The elements of the square root of the imaginary
part of the effective tensors $\sqrt{{\bar{\bar{\boldsymbol\varepsilon }}}_{\rm eff}^{I}\left( \omega  \right)}$ and  $\sqrt{{\bar{\bar{\boldsymbol\mu }}}_{\rm eff}^{-1\,I}\left( \omega  \right)}$ can be expressed as:
\begin{subequations}
\begin{eqnarray}
e_{11}&=&\sqrt{\bar{\bar\varepsilon}_{\rm eff ,xx}^{I}},\\
e_{22}&=&\frac{(-\bar{\bar\varepsilon}_{\rm eff ,yy}^{I}+\bar{\bar\varepsilon}_{\rm eff ,zz}^{I})({\cal M}_e-{\cal N}_e) }{2{\cal D}_e}\\
&&+\frac{ {\cal D}_e({\cal M}_e+{\cal N}_e)}{2{\cal D}_e},\nonumber\\
e_{23}&=&\frac{\Big((\bar{\bar\varepsilon}_{\rm eff ,yy}^{I}-\bar{\bar\varepsilon}_{\rm eff ,zz}^{I})^2-{\cal D}_e^2\Big)({\cal M}_e-{\cal N}_e) }{2i\bar{\bar\varepsilon}_{\rm eff ,zy}^{*}{\cal D}_e},\\
e_{32}&=&\frac{-i\bar{\bar\varepsilon}_{\rm eff ,zy}^{*}({\cal M}_e-{\cal N}_e)}{2{\cal D}_e},\\
e_{33}&=&\frac{(\bar{\bar\varepsilon}_{\rm eff ,yy}^{I}-\bar{\bar\varepsilon}_{\rm eff ,zz}^{I})({\cal M}_e-{\cal N}_e) }{2{\cal D}_e}\\
&&+\frac{{\cal D}_e({\cal M}_e+{\cal N}_e)}{2{\cal D}_e},\nonumber\\
m_{11}&=&\sqrt{\bar{\bar\mu}_{\rm eff ,xx}^{-1\,I}},\\
m_{22}&=&\frac{(-\bar{\bar\mu}_{\rm eff ,yy}^{-1\,I}+\bar{\bar\mu}_{\rm eff ,zz}^{-1\,I})({\cal M}_m-{\cal N}_m) }{2{\cal D}_m}\\
&&+\frac{{\cal D}_m({\cal M}_m+{\cal N}_m)}{2{\cal D}_m},\nonumber\\
m_{23}&=&\frac{\Big((\bar{\bar\mu}_{\rm eff ,yy}^{-1\,I}-\bar{\bar\mu}_{\rm eff ,zz}^{-1\,I})^2-{\cal D}^2_m\Big)({\cal M}_m-{\cal N}_m) }{2i\bar{\bar\mu}_{\rm eff ,zy}^{-1\,*}{\cal D}_m},\,\,\,\,\,\\
m_{32}&=&\frac{-i\bar{\bar\mu}_{\rm eff ,zy}^{-1\,*}({\cal M}_m-{\cal N}_m)}{2{\cal D}_m},\\
m_{33}&=&\frac{(\bar{\bar\mu}_{\rm eff ,yy}^{-1\,I}-\bar{\bar\mu}_{\rm eff ,zz}^{-1\,I})({\cal M}_m-{\cal N}_m) }{2{\cal D}_m}\\
&&+\frac{{\cal D}_m({\cal M}_m+{\cal N}_m)}{2{\cal D}_m},\nonumber
\end{eqnarray}
\end{subequations}
where ${\cal D}_e= \sqrt{(\bar{\bar\varepsilon}_{\rm eff ,yy}^{I}-\bar{\bar\varepsilon}_{\rm eff ,zz}^{I})^2+|\bar{\bar\varepsilon}_{\rm eff ,zy}^{I}|^2}$, ${\cal M}_e=\sqrt{(\bar{\bar\varepsilon}_{\rm eff ,yy}^{I}+\bar{\bar\varepsilon}_{\rm eff ,zz}^{I}-{\cal D}_e)/2}$ and ${\cal N}_e=\sqrt{(\bar{\bar\varepsilon}_{\rm eff ,yy}^{I}+\bar{\bar\varepsilon}_{\rm eff ,zz}^{I}+{\cal D}_e)/2}$. Furthermore, the parameters ${\cal D}_m $, ${\cal M}_m $ and ${\cal N}_m $ are obtained from the previous relations by replacing $\bar{\bar\varepsilon}_{\rm eff}^{I}$ with $\bar{\bar\mu}_{\rm eff}^{-1\,I}$.
%
\section{Boundary Conditions }\label{App:Boundary Conditions}
Considering the continuity of the tangential electric and magnetic
fields across the boundaries of the MDS, and taking into account that the vector potential with components~(\ref{vector potential components x}) and~(\ref{vector potential components y}) is continuously differentiable at the interface between the MDS and the vacuum, the boundary conditions at  $z=z_j\,\,(j=1,2)$ can be written as:
\begin{subequations}\label{tangential components Ex,Ey}
\begin{eqnarray}
&& \sqrt{{{\xi }_{j+1}}}\frac{{{\mu }_{{\rm eff\,{j+1} , yy} }}}{{{n}_{j+1}}}\left[ {{e}^{{i {\eta_{ j+1}} \omega {{z}_{j}}}/{c} }}\hat{a}_{x+}^{\left( j+1 \right)}\left(
  {{z}_{j}},\omega  \right)\right.\nonumber\\
&& \hspace{2.5cm} +\left.{{e}^{{-i {\eta_{j+1}}\,\omega {{z}_{j}}}/{c} }}\hat{a}_{x-}^{\left( j+1 \right)}\left( {{z}_{j}},\omega  \right) \right]
  \nonumber\\
&=&\sqrt{{{\xi }_{j}}}\frac{{{\mu }_{{\rm eff \,j, yy}}}}{{{n}_{j}}}\left[ {{e}^{{i{\eta_{j\,}}\omega {{z}_{j}}}/{c} }}\hat{a}_{x+}^{\left( j
 \right)}\left( {{z}_{j}},\omega  \right)\right.\\
&& \hspace{2.5cm}  +\left.{{e}^{{-i{\eta_{j}} \omega {{z}_{j}}}/{c} }}\hat{a}_{x-}^{\left( j \right)}\left( {{z}_{j}},\omega  \right)
 \right],\nonumber
\\
\nonumber\\
\nonumber\\
&& \sqrt{{{{{\xi }'}}_{j+1}}}\frac{{ {{\mu }_{\rm eff\,xx,j+1}} }}{{{n}_{j+1}}}\left[ {{e}^{{i{\eta_{ j+1}} \omega
 {{z}_{j}}}/{c} }}\hat{a}_{y+}^{\left( j+1 \right)}\left( {{z}_{j}},\omega  \right)\right.\nonumber\\
&& \hspace{2.5cm} \left. +{{e}^{{-i{\eta_{j+1}}\omega {{z}_{j}}}/{c} }}\hat{a}_{y-}^{\left( j+1
 \right)}\left( {{z}_{j}},\omega  \right) \right] \nonumber \\
&=&\sqrt{{{{{\xi }'}}_{j}}}\frac{{ {{\mu }_{\rm eff\,xx,j}}}}{{{n}_{j}}}\left[ {{e}^{{i{\eta_{ j}}\,\omega
 {{z}_{j}}}/{c} }}\hat{a}_{y+}^{\left( j \right)}\left( {{z}_{j}},\omega  \right)\right.\\
 && \hspace{2.5cm} \left.+{{e}^{{-i{\eta_{j}} \omega {{z}_{j}}}/{c} }}\hat{a}_{y-}^{\left( j
 \right)}\left( {{z}_{j}},\omega  \right) \right]. \nonumber
\end{eqnarray}
\end{subequations}
and
\begin{subequations}\label{tangential components Hx,Hy}
\begin{eqnarray}
&&\sqrt{{{\xi }_{j+1}}}\left[ {{e}^{{i {\eta_{j+1}}\omega {{z}_{j}}}/{c} }}\hat{a}_{x+}^{\left( j+1 \right)}\left( {{z}_{j}},\omega
  \right)\right.\nonumber\\
&&\hspace{1.2cm}\left.-{{e}^{{-i {\eta_{j+1}}\,\omega {{z}_{j}}}/{c} }}\hat{a}_{x-}^{\left( j+1 \right)}\left( {{z}_{j}},\omega  \right) \right]  \\
&=& \sqrt{{{\xi }_{j}}}\left[ {{e}^{{i {\eta_{j}} \omega {{z}_{j}}}/{c} }}\hat{a}_{x+}^{\left( j \right)}\left(
 {{z}_{j}},\omega  \right)-{{e}^{{-i\,{\eta_{ j }}\omega z}/{c} }}\hat{a}_{x-}^{\left( j \right)}\left( {{z}_{j}},\omega  \right) \right] \nonumber\\
\nonumber \\
\nonumber \\
&& \sqrt{{{{{\xi }'}}_{j+1}}}\left[ {{e}^{{i {\eta_{\,j+1}} \omega {{z}_{j}}}/{c} }}\hat{a}_{y+}^{\left( j+1 \right)}\left( {{z}_{j}},\omega
   \right)\right.\nonumber\\
&&\hspace{1.2cm}\left.-{{e}^{{-i {\eta_{ j+1}} \omega {{z}_{j}}}/{c} }}\hat{a}_{y-}^{\left( j+1 \right)}\left( {{z}_{j}},\omega  \right) \right] \\
& =&\sqrt{{{{{\xi }'}}_{j}}}\left[ {{e}^{{i {\eta_{j}} \omega {{z}_{j}}}/{c} }}\hat{a}_{y+}^{\left( j \right)}\left(
 {{z}_{j}},\omega  \right)-{{e}^{{-i\,{\eta_{j}} \omega {{z}_{j}}}/{c} }}\hat{a}_{y-}^{\left( j \right)}\left( {{z}_{j}},\omega  \right) \right],\,\,\,\,\,\,\,\,\nonumber
\end{eqnarray}
\end{subequations}
where,  $\eta_j $ and $\kappa_j $ are, respectively, the real and imaginary parts of the refractive index $n_j$. Here, $n_2=n_{\rm eff}$  is the refractive index of the moving MDS in the laboratory frame. Notice that because of the moving MDS is surrounded by vacuum, we have $\varepsilon_1=\varepsilon_3=\mu_1=\mu_3=1$. Therefore, we have: $\eta_1=\eta_3=1$, $\kappa_1=\kappa_3=0$, $\mu_{\rm eff\,1,xx}=\mu_{\rm eff\,1,yy}=\mu_{\rm eff\,3,xx}=\mu_{\rm eff\,3,yy}=1$ and $\varepsilon_{\rm eff\,1,xx}=\varepsilon_{\rm eff\,1,yy}=\varepsilon_{\rm eff\,3,xx}=\varepsilon_{\rm eff\,3,yy}=1$.

\section{Elements of Transformation Matrix }\label{App:Components of Transformation Matrix}
We can now relate the operators $\hat{a}_{\sigma \pm }^{(j+1)}\left( {{z}_{j}},\omega  \right)$ and $\hat{a}_{\sigma \pm}^{(j)}\left( {{z}_{j}},\omega  \right)$ to each other using the boundary conditions~(\ref{tangential components Ex,Ey}) and~(\ref{tangential components Hx,Hy}).
After straightforward calculations, we arrive at Eq.~(\ref{operator relation in neighboring layers}), where the elements of the transformation matrix ${\mathbb S}^{(j)}$ read as:
\begin{subequations}\label{transformation matrix 11x}
\begin{eqnarray}
{\mathbb S}_{11,x}^{\left( j \right)}&=&\frac{\sqrt{{{\xi }_{j}}}}{\sqrt{{{\xi }_{j+1}}}}\frac{{{n}_{j+1}}{{\mu }_{\rm eff\,j,yy}}+{{n}_{j}}{{\mu }_{\rm
eff\,j+1,yy}}}{2{{n}_{j}}{{\mu }_{\rm eff\,j+1,yy}}}\nonumber\\
&&\times {{e}^{-i \left( {\eta_{ j+1}}- {\eta_{ j}} \right){{z}_{j}}{\omega }/{c} }} \nonumber\\
&=& {\mathbb S}_{22,x}^{\left( j \right)}{{e}^{-i\,2\,\left( {\eta_{j+1}}- {\eta_{j}} \right){{z}_{j}}{\omega }/{c} }},\\
{\mathbb S}_{12,x}^{\left( j \right)}&=&\frac{\sqrt{{{\xi }_{j}}}}{\sqrt{{{\xi }_{j+1}}}}\frac{{{n}_{j+1}}{{\mu }_{\rm eff\,j,yy}}-{{n}_{j}}{{\mu }_{\rm
eff\,j+1,yy}}}{2{{n}_{j}}{{\mu }_{\rm eff\,j+1,yy}}}\nonumber\\
&&\times{{e}^{-i\left( {\eta_{j+1}}+ {\eta_{ j}} \right){{z}_{j}}{\omega }/{c} }}\nonumber\\
 &=&{\mathbb S}_{21,x}^{\left( j \right)}{{e}^{-i 2\left( {\eta_{j+1}}+ {{\beta }_{j}} \right){{z}_{j}}{\omega }/{c} }},\\
{\mathbb S}_{11,y}^{\left( j \right)}&=&\frac{\sqrt{{{{{\xi }'}}_{j}}}}{\sqrt{{{{{\xi }'}}_{j+1}}}}\frac{{{n}_{j+1}}{{\mu }_{\rm eff\,j,xx}}+{{n}_{j}}{{\mu }_{\rm
eff\,j+1,xx}}}{2{{n}_{j}}{{\mu }_{\rm eff\,j+1,xx}}}\nonumber\\
&&\times{{e}^{-i\left( {\eta_{j+1}}- {\eta_{j}} \right){{z}_{j}}{\omega }/{c} }}\nonumber\\
  &=&{\mathbb S}_{22,y}^{\left( j \right)}{{e}^{-i 2\left( {\eta_{ j+1}}- {\eta_{ j}} \right){{z}_{j}}{\omega }/{c} }},\\
{{\mathbb S}_{12,y}}&=&\frac{\sqrt{{{{{\xi }'}}_{j}}}}{\sqrt{{{{{\xi }'}}_{j+1}}}}\frac{{{n}_{j+1}}{{\mu }_{\rm eff\,j,xx}}-{{n}_{j}}{{\mu }_{\rm
eff\,j+1,xx}}}{2{{n}_{e,j}}{{\mu }_{\rm eff\,j+1,xx}}}\nonumber\\
&&\times{{e}^{-i\left( {\eta_{j+1}}+ {\eta_{ j}} \right){{z}_{j}}{\omega }/{c} }}\nonumber\\
  &=&{{\mathbb S}_{21,y}}{{e}^{-i 2\left( {\eta_{j+1}}+ {\eta_{j}} \right){{z}_{j}}{\omega }/{c} }}.
\end{eqnarray}
\end{subequations}

%
\section{Elements of Absorption Matrix }\label{App:Components of Absorbing Matrix}
The elements of the characteristic absorption matrix $\mathbb{A}_{\sigma }$ are expressed as follows:
\begin{subequations}\label{absorbing matrix}
\begin{eqnarray}
\mathbb{A}_{11,x}&=&\sqrt{{\kappa}{{\xi } } {{c}_{x+}}}{{t}_{12,x}}\,\vartheta {{e}^{-i\omega {l}/{2c} }}\left( 1+{{e}^{i{{n}_{\rm eff}}\omega
{l}/{c} }}{{r}_{23,x}} \right),\\
\mathbb{A}_{12,x}&=&\sqrt{{\kappa}{{\xi } }{{c}_{x-}}}{{t}_{12,x}}\vartheta  {{e}^{-i \omega {l}/{2c} }}\left(
1-{{e}^{i{{n}_{\rm eff}}\omega {l}/{c} }}{{r}_{23,x}} \right),\\
\mathbb{A}_{21,x}&=&\sqrt{{\kappa } {{\xi } } {{c}_{x+}}}{{t}_{32,x}}\vartheta  {{e}^{-i \omega {l}/{2c} }} \left(
1+{{e}^{i{{n}_{\rm eff}}\omega {l}/{c} }}{{r}_{21,x}} \right),\\
\mathbb{A}_{22,x}&=&\sqrt{{\kappa }{{\xi } } {{c}_{x-}}}{{t}_{32,x}}\vartheta   {{e}^{-i \omega {l}/{2c} }} \left(
{{e}^{i{{n}_{\rm eff}}\,\omega {l}/{c} }}{{r}_{21,x}}-1 \right),\\
\mathbb{A}_{11,y}&=&\sqrt{{\kappa }{{{{\xi }'}} } {{c}_{y+}}} {{t}_{12,y}} {\vartheta }' {{e}^{-i \omega {l}/{2c} }} \left(
1+{{e}^{i{{n}_{\rm eff}}\omega {l}/{c} }}{{r}_{23,y}} \right),\\
\mathbb{A}_{12,y}&=&\sqrt{{\kappa }{{{{\xi }'}} }{{c}_{y-}}}{{t}_{12,y}} {\vartheta }' {{e}^{-i \omega {l}/{2c} }} \left(
1-{{e}^{i{{n}_{\rm eff}}\,\omega {l}/{c} }}{{r}_{23,y}} \right),\\
\mathbb{A}_{21,y}&=&\sqrt{{\kappa}{{{{\xi }'}} } {{c}_{y+}}}{{t}_{32,y}}{\vartheta }' {{e}^{-i \omega {l}/{2c} }}\left(
1+{{e}^{i{{n}_{\rm eff}}\omega {l}/{c} }}{{r}_{21,y}} \right),\\
\mathbb{A}_{22,y}&=&\sqrt{{\kappa} {{{{\xi }'}} } {{c}_{y-}}}{{t}_{32,y}}{\vartheta }'  {{e}^{-i \omega {l}/{2c} }}\left(
{{e}^{i{{n}_{\rm eff}}\omega {l}/{c} }}{{r}_{21,y}}-1 \right).\,\,\,\,\,\,\,\,\,\,\,\,\,\,
\end{eqnarray}
\end{subequations}
where
\begin{subequations}
\begin{eqnarray}
{\vartheta }'&=&{{\left[ 1-r_{21,y}^{2}{{e}^{2i{{n}_{\rm eff}}\omega {l}/{c}}} \right]}^{-1}},\\
\vartheta &=&{{\left[ 1-r_{21,x}^{2}{{e}^{2i{{n}_{\rm eff}}\omega
{l}/{c}}} \right]}^{-1}},\\
\end{eqnarray}
\end{subequations}
and in which
\begin{subequations}
\begin{eqnarray}
{{r}_{12,x}}&=&{{r}_{32,x}}=-{{r}_{21,x}}=-{{r}_{23,x}}=\frac{{{\mu }_{\rm eff ,yy }}-{{n}_{\rm eff}}}{{{\mu }_{\rm eff ,yy }}+{{n}_{\rm eff}}},\,\,\,\,\,\,\,\,\\
{{t}_{21,x}}&=&{{t}_{23,x}}=\frac{2{{n}_{\rm eff}}}{{{\mu }_{\rm eff ,yy }}+{{n}_{\rm eff}}},\\
{{r}_{12,y}}&=&{{r}_{32,y}}=-{{r}_{21,y}}=-{{r}_{23,y}}=\frac{{{\mu }_{\rm eff ,xx }}-{{n}_{\rm eff}}}{{{\mu }_{\rm eff ,xx }}+{{n}_{\rm eff}}},\\
{{t}_{12,y}}&=&{{t}_{32,y}}=\frac{2{{\mu }_{\rm eff ,xx }}}{{{\mu }_{\rm eff ,xx }}+{{n}_{\rm eff}}},\\
{{t}_{21,y}}&=&{{t}_{23,y}}=\frac{2{{n}_{\rm eff}}}{{{\mu }_{\rm eff ,xx }}+{{n}_{\rm eff}}}.\\
{{t}_{12,x}}&=&{{t}_{32,x}}=\frac{2{{\mu }_{\rm eff ,yy }}}{{{\mu }_{\rm eff ,yy }}+{{n}_{\rm eff}}}.
\end{eqnarray}
\end{subequations}

%
%


\begin{thebibliography}{10}
%
\bibitem{Minkowski1908}	H. Minkowski, Nachr. Ges. Wiss. Gotingen 53, (1908).
%
\bibitem{Sommerfeld1964} A. Sommerfeld, "Electrodynamics", (Academic, New York, 1964).
%
\bibitem{Landau1984} L. D. Landau, E. M. Lifshitz and L. P. Pitaevskii, "Electrodynamics of Continuous Media" (Pergamon, Oxford,1984).
%
\bibitem{Gordon1923} W. Gordon, Ann. Phys. (Leipzig) 72, 421 (1923).
%
\bibitem{Leonhardt1999} U. Leonhardt, and P. Piwnicki, Phys. Rev. Lett. 60, 4301 (1999).
%
\bibitem{Carusotto2001} I. Carusotto, M. Artoni, G.C. La Rocca, F. Bassani, Phys. Rev. Lett. 86, 2549 (2001).
%
\bibitem{Carusotto2003} I. Carusotto, M. Artoni, G.C. La Rocca, F. Bassani, Phys. Rev. A 68, 063819 (2003).
%
\bibitem{Strekalov2004} D. Strekalov, A.B. Matsko. N. Yu., L. Maleki,
Phys. Rev. Lett. 93, 023601 (2004)
%
\bibitem{Nag1956} B. D. Nag, and A. M. Sayied, Proc. R. Soc. London, Ser. A, 235, 544 (1956).
%
\bibitem{Kong1975} J. A. Kong, Theory of Electromagnetic Waves (Wiley, New York, 1975).
%
\bibitem{Pauli1958} W. Pauli, TIzeory of Relativity [Pergamon, New York, 1958).
%
\bibitem{Sommerfeld1959} A. Sommerfeld, Opfik, 2nd ed. (Akademische, Leipzig, 1959).
%
\bibitem{Yeh1965} C. Yeh, J. Appl. Phys. 36, 3513 (1965).
%
\bibitem{Yeh1966a} C. Yeh, J. Appl. Phys. 37, 3079 (1966).
%
\bibitem{Yeh1966b} C. Yeh and K. F. Casey, Phys. Rev. 144, 665 (1966).
%
\bibitem{Payati1967} V. P. Payati, J. Appl. Phys. 38, 652 (1967).
%
\bibitem{Shiozawa1967} T. K. Shiozawa, K. Hazawa, and N. Kumagai, J. Appl. Phys. 38, 4459
(1967).
%
\bibitem{Kong1968} J. A. Kong and D. K. Cheng, J. Appl. Phys. 39, 2282 (1968).
%
\bibitem{Shiozawa1972} T. Shiozawa, and S. Seikai, IEEE Trans. Antennas Propag, 20, 455 (1972).
%
%
\bibitem{Huang1994} Y. X. Huang, J. Appl. Phys. 76, 2575  (1994).
%
\bibitem{Leonhardt2000} U. Leonhardt, Nature (London) 415, 406 (2000).
%
\bibitem{Fiurasek2002} J. Fiurasek, U. Leonhardt, and R. Parentani, Phys. Rev. A 65, 011802 (2002).
%
\bibitem{Cook1995} R. J. Cook, H. Fearn and P. W. Milonni, Am. J. Phys. 63, 705 (1995).

\bibitem{Grzegorczyk2006} T. M. Grzegorczyk, J. A. Kong, Phys. Rev. B 74, 033102 (2006).
%
\bibitem{Da2007} H. X. Da, Z. Y. Li, Phys. Rev. B 76, 012409 (2007).
%
\bibitem{Fumeron2011} S. Fumeron, P. Vaveliuk, E. Faudot, and F. Moraes, J. Opt. Soc. Am. B 28, 765 (2011).
%
\bibitem{Lin2016} Sh. Lin, R. Zhang, Y. Zhai, J. Wei, Q. Zhao, J. Opt. 18, 085603 (2016).
%
\bibitem{Svidzinsky2017} A. A. Svidzinsky, F. Li, and X. Zhang, Phys. Rev. Lett. 118, 012105 (2017).
%
\bibitem{Dodonov2010} V. V. Dodonov, Phys. Scr. 82, 038105 (2010).
%
\bibitem{Silveirinha2014} M. G. Silveirinha, Phys. Rev. X 4, 031013 (2014).
%
\bibitem{Fulling1976} S. A. Fulling and P. C.W. Davies, Proc. R. Soc. A 348, 393 (1976).
%
\bibitem{Barton1996} G. Barton, Ann. Phys. (N.Y.) 245, 361 (1996).
%
\bibitem{Volokitin2007} A. I. Volokitin and B. N. J. Persson, Rev. Mod. Phys. 79, 1291 (2007).
%
\bibitem{Maghrebi2012} M. F. Maghrebi, R. L. Jaffe, and M. Kardar, Phys. Rev. Lett. 108, 230403 (2012).
%
\bibitem{Maghrebi2013} M. F. Maghrebi, R. Golestanian, and M. Kardar, Phys. Rev. D, 87, 025016 (2013).
%
\bibitem{Manjavacas2010} A. Manjavacas and F. J. Garc\'{\i}a de Abajo, Phys. Rev. Lett. 105, 113601 (2010).
%
\bibitem{Zhao2012} R. Zhao, A. Manjavacas, F. J. Garc\'{\i}a de Abajo, J. B. Pendry, Phys. Rev. Lett. 109, 123604 (2012).
%
\bibitem{Gruner1996} T. Gruner and D.-G. Welsch, Phys. Rev. A 54, 1661 (1996).
%
\bibitem{Artoni1997} M. Artoni, R. Loudon, Phys. Rev. A 55, 1347 (1997).
%
\bibitem{Artoni1998a} M. Artoni, R. Loudon, Phys. Rev. A 59, 2279 (1998).
%
\bibitem{Artoni1998b} M. Artoni, R. Loudon, Phys. Rev. A 57, 622 (1998).
%
\bibitem{Matloob2000} R. Matloob, G. Pooseh, Opt. Commun. 181, 109 (2000).
%
\bibitem{Khanbekyan2003} M. Khanbekyan, L. Kn\"{o}ll, and D.-G. Welsch, Phys. Rev. A 67, 063812 (2003).
%
\bibitem{Amooghorban2013} E. Amooghorban, N.A. Mortensen, M. Wubs, Phys. Rev. Lett. 110, 153602 (2013).
%
\bibitem{Amooghorban arXiv} E. Amooghorban, M. Wubs, arXiv:1606.07912.
%
\bibitem{Amooghorban2014} E. Amooghorban and A. Mahdifar, Iran. J. Phys. Res. 14, 7 (2014).
%
\bibitem{Aghbolaghi2017} A. Aghbolaghi, E. Amooghorban, and A. Mahdifar, Eur. Phys. J. D 71, 272 (2017).
%
\bibitem{Matloob2005a} R. Matloob, Phys. Rev. A 71, 062105 (2005).
%
\bibitem{Matloob2005b} R. Matloob, Phys. Rev. A 72, 062103 (2005).
%
\bibitem{Amooshahi2009} M. Amooshahi, Phys. Rev. A 54, 115 (2009).
%
\bibitem{Kheirandish2011} F. Kheirandish, and S. Salimi, Phys. Rev. A 84, 062122 (2011)
%
\bibitem{Horsley2012} S. A. R. Horsley, Phys. Rev. A 86, 023830 (2012).
%
\bibitem{Chen1983} H. C. Chen, Theory of Electromagnetic Waves (McGraw-Hill, New York, 1983).
%
\bibitem{Matloob1996} R. Matloob, and R. Loudon, Phys. Rev. A 53, 4567 (1996).
%
\bibitem{Knoll2001} L. Knoll, S. Scheel and D.-G. Welsch, in Coherence and Statistics
of Photons and Atoms, edited by J. Perina (Wiley, New York, 2001).
%
\bibitem{Dong2011} Y. Dong, and X. Zhang, J. Opt. 13, 035401 (2011).
%
\bibitem{Hoseinzadeh2017} M. Hoseinzadeh, E. Amooghorban, A. Mahdifar, Iran. J. Phys. Res. 16, 305 (2017).
%
\bibitem{Scully1997} M. O. Scully and M. S. Zubairy, "Quantum Optics", (Cambridge University Press, Cambridge, England, 1997).



%
%
%


\end{thebibliography}
\end{document}